\documentclass[a4paper,fleqn]{cas-sc}

\usepackage[authoryear]{natbib}

\usepackage[capitalise]{cleveref}
\usepackage{booktabs}
\usepackage{siunitx}
\usepackage{xurl}
\usepackage{upgreek}
\usepackage{bm}

\def\tsc#1{\csdef{#1}{\textsc{\lowercase{#1}}\xspace}}
\tsc{WGM}
\tsc{QE}

\begin{document}
\let\WriteBookmarks\relax
\def\floatpagepagefraction{1}
\def\textpagefraction{.001}

\shorttitle{Hydrodynamic drag load on side-by-side flexible blades}

\shortauthors{Z. Wei, \textit{et~al.}}

\title[mode=title]{Experimental and theoretical investigation of drag loads on side-by-side flexible blades in a uniform current}



%

\author[DTU,NTNU]{Zhilong Wei}[]





\credit{Conceptualization, Methodology, Software, Validation, Formal analysis, Writing - Original Draft, Visualization}

\author[NTNU]{Trygve Kristiansen}[]





\credit{Methodology, Investigation, Writing - Review \& Editing, Supervision, Funding acquisition}

\author[NTNU]{David Kristiansen}[]





\credit{Methodology, Investigation, Writing - Review \& Editing, Supervision, Funding acquisition}

\author[DTU]{Yanlin Shao}[orcid=0000-0002-9080-8438]

\cormark[1]


\ead{yshao@dtu.dk}


\credit{Conceptualization, Methodology, Investigation, Writing - Review \& Editing, Supervision, Project administration, Funding acquisition}

\affiliation[DTU]{organization={Department of Civil and Mechanical Engineering, Technical University of Denmark},
            city={Kgs. Lyngby},
            postcode={2800},
            country={Denmark}}

\affiliation[NTNU]{organization={Department of Marine Technology, Norwegian University of Science and Technology},
            city={Trondheim},
            postcode={7491},
            country={Norway}}







\cortext[1]{Corresponding author}



\begin{abstract}
This study investigates the hydrodynamic drag loads on side-by-side flexible blades in a uniform current through experimental and theoretical approaches.
Four silicone rubber blade models with varying dimensions were arranged side by side in aggregates and tested in a circulating water tunnel.
The experiments cover the static regime and the flutter regime.
We examine four governing non-dimensional parameters to assess their effects on the bulk drag coefficient $C_{D,\mathrm{bulk}}$ and the onset of flutter: the drag-to-stiffness ratio $\mathrm{Ca}$, the buoyancy-to-stiffness ratio $\mathrm{B}$, the mass ratio of fluid inertia to total system inertia $\upbeta$, and the slenderness parameter $\uplambda$.
The results show that in the static regime $C_{D,\mathrm{bulk}}$ decreases with increasing $\mathrm{Ca}$ at a rate closely related to $\mathrm{B}$ starting at $\mathrm{Ca}/\mathrm{B} > O(1)$, and settles to an almost constant value in the flutter regime.
Increasing $\upbeta$, $\mathrm{B}$, or $\uplambda$ delays the onset of flutter.
By introducing the equivalent thickness and bending stiffness, existing theoretical models for individual blades are utilized to predict drag reduction of side-by-side blade aggregates.
The analytical model accurately predicts drag reduction in the static regime, while the numerical model predicts the onset of flutter and drag reduction in both regimes when appropriate cross-sectional hydrodynamic coefficients are applied.
Meanwhile, we investigate the reactive force model term by term to identify their impact on system stability and drag reduction, demonstrating its suitability for highly compliant blades in uniform flows.
\end{abstract}

\begin{keywords}
flexible blades \sep drag reduction \sep reconfiguration \sep bulk drag coefficient \sep flutter 
\end{keywords}

\maketitle

\section{Introduction}\label{sec:intro}
\begin{figure}
    \centerline{\includegraphics[height=0.25\linewidth]{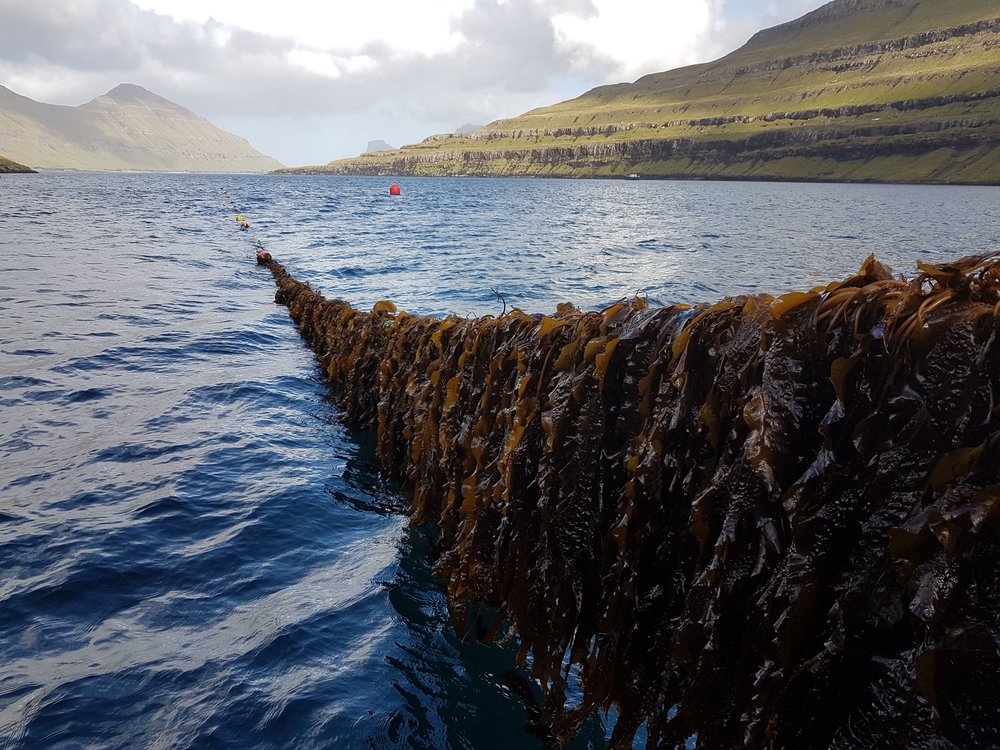}
    \includegraphics[height=0.25\linewidth]{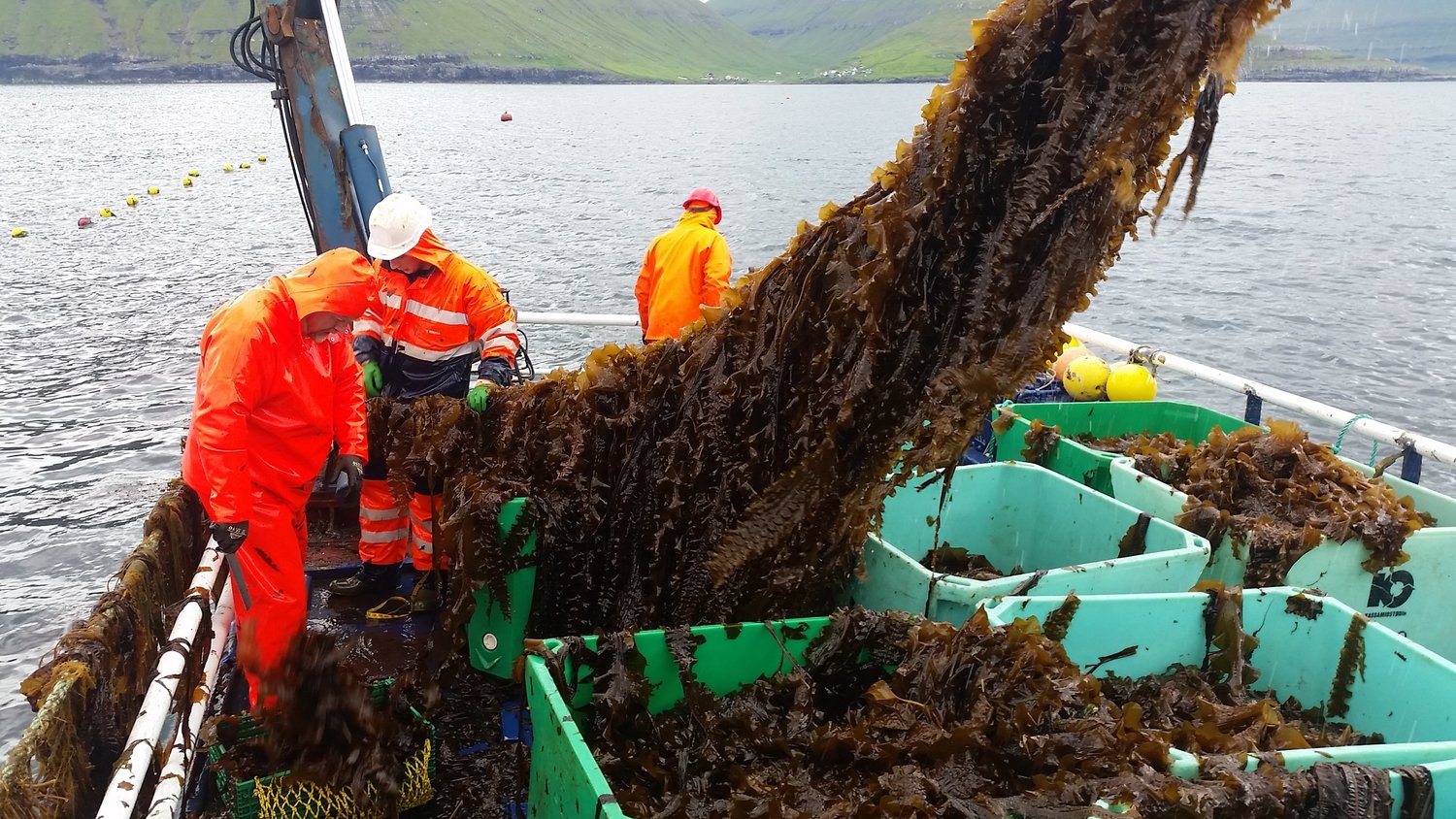}}
    \caption{Cultivated seaweed (left) and seaweed harvest (right) in the Faroe Islands. Photo courtesy of Ocean Rainforest.}
    \label{fig:seaweed_cultivation}
\end{figure}

Flexible structures in nature and industry, from plant leaves to sails, exhibit compliant behavior and response to aero-hydrodynamic loads different from rigid bodies.
Flexible structures subjected to transverse flows are pushed to align with the flow direction, and the drag they experience is modified by the deformation.
\citet{Vogel1996} suggested the use of the term \textit{reconfiguration} when dealing with biological structures to emphasize the strategy of being flexible to reduce the drag.
Seaweeds are one of the most common examples.
In the case of cultivated seaweeds, as shown in \cref{fig:seaweed_cultivation}, to maximize the yield, the seaweed blades usually grow densely side by side on longlines~\citep{Grebe2019,Campbell2019,Zhu2021} instead of being separated from each other.

The theory of static reconfiguration and drag reduction of an individual flexible blade in a steady background flow is well established.
\citet{Gosselin2010} introduced the reconfiguration number to quantify drag reduction in their wind tunnel experiments on flexible rectangular plates and cut disks.
They developed a simple analytical model based on a force balance between drag and bending stiffness, which predicted well the observed drag reduction.
\citet{Luhar2011} incorporated gravity and buoyancy and used the effective length to represent drag reduction.
Both concepts (reconfiguration number and effective length) have been widely used since then, for example, the former in \citet{Henriquez2014}, \citet{Whittaker2015} and \citet{Baskaran2023}, and the latter in \citet{Lei2021} and \citet{Sun2024}.
However, the relationship between them has not been clarified.

Flexible blades exposed to a transverse flow tend to deform themselves to align with the flow direction.
Beyond a certain flow speed, flutter, which is a dynamic instability resulting from the competition between reactive fluid forces and structural stiffness~\citep{Leclercq2018a}, starts to occur.
The analytical models based on the static assumptions mentioned above are no longer applicable in this dynamic regime.
\citet{Leclercq2018a} developed a numerical model to study drag reduction in the flutter regime and concluded that
after flutter occurs, the drag is not reduced as rapidly as in the static regime.
\citet{Boukor2024} reached a similar conclusion from their experiments, referring to this phenomenon as the flutter limitation of drag reduction.
It is unknown whether this limitation exists in the scenario of side-by-side blades and, if it does, how this limitation differs from the individual blade scenario.
In the analytical and numerical modeling of flutter in cross flow (which denotes flow perpendicular to the blade width axis throughout this work), such as \citet{Leclercq2018a}, the resistive drag~\citep{Taylor1952} and the reactive force~\citep{Lighthill1971,Candelier2011} were used as the hydrodynamic load model.
However, the precise roles of the terms in the hydrodynamic load model in triggering flutter and in stabilizing or destabilizing the system remain unclear.

Studies on the hydrodynamic drag of side-by-side flexible blade aggregates are relatively scarce in the literature.
\citet{Buck2005} conducted towing tests on individual blades of wild and cultivated \textit{Laminaria saccharina}, as well as on aggregates of laminarians.
For cultivated algae from an exposed offshore farm, the kelp blades were flat and narrow, and the reported bulk drag coefficient on the cultivated algae initially decreased with increasing current velocity before reaching some constant values with the current velocity, as shown in their Figure 7.
They also found that drag coefficients for ruffled and wide specimens from sheltered environments were 2--4 times larger than those for flat and narrow farmed specimens from exposed areas.
\citet{Fredriksson2020} performed towing tests on a dense aggregate of artificial kelp blades and reported drag-area values.
\citet{Lei2021} observed that the drag increased with velocity at a rate less than quadratic in towing tests on dense aggregates of cultivated \textit{Saccharina latissima}.
However, the existing experimental studies do not present a systematic analysis utilizing governing non-dimensional parameters to unravel the physics.

When multiple flexible blades are arranged in aggregates, their interactions with the flow and with each other introduce additional complexities absent in isolated blades.
Studies have found that the drag on an aggregate is less than the sum of the individual drag forces~\citep{Carrington1990,Johnson2001,Buck2005}.
The relationship between the drag on an aggregate and that on an individual blade is not clearly understood, let alone the drag on side-by-side aggregates.
As mentioned earlier, well-developed theoretical models exist for an individual blade in an infinite fluid domain. If these models can be moderately modified to study the drag load on side-by-side blade aggregates, they could provide an efficient alternative to developing complex three-dimensional models from scratch.

The first objective of this paper is to extract the bulk drag coefficients for side-by-side blades in a current, using comprehensive experimental data that spans both the static and flutter regimes.
Leveraging this data set, the study reexamines key non-dimensional parameters to analyze their influence on the bulk drag coefficient in both regimes and their roles on the onset of flutter. Additionally, the paper aims to assess whether established analytical and numerical models for individual blades can adequately capture drag loads in side-by-side blade configurations. Furthermore, the reactive force model~\citep{Lighthill1971,Candelier2011} will be scrutinized term by term to evaluate its impact on drag reduction and system stability.

The paper is organized as follows.
In \cref{sec:analytical_model}, we present the dynamic and static governing equations of motion of flexible blades in a steady current, and their similarity to the classic small-amplitude flutter equation for an undamped beam in axial flow.
The experimental design and setup are described in \cref{sec:expriments}, followed by an overview of the numerical model and a convergence study in \cref{sec:numerical_model}.
Experimental, analytical, and numerical results are compared and discussed in \cref{sec:results}.
In \cref{sec:discussion}, we examine the hydrodynamic load model, particularly the reactive force, and assess the contributions of individual terms to drag reduction and system stability. We also discuss the limitations of the theoretical models and the experimental approach in this section.
Finally, conclusions are provided in \cref{sec:conclusion}.

\section{Theoretical description}\label{sec:analytical_model}

\subsection{Governing equations}\label{sec:governing_equations}
In this section, we first consider the physical problem of an individual blade in an infinite fluid domain and then extend this model for an aggregate of multiple blades.

\begin{figure}
    \centerline{\includegraphics{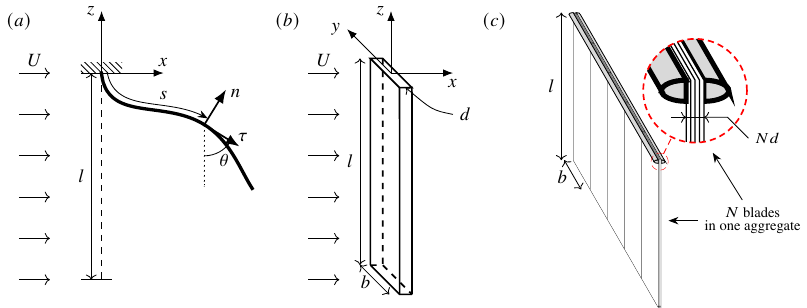}}
    \caption{(a) Side view of the deformed blade subjected to a uniform flow $U$ in the $x$-direction. The length of the blade is $l$, and $s$ is the arc length measured from the base of the blade. The local angle $\theta$ is the angle between the tangential direction and the negative $z$-axis. The local tangential vector and the normal vector are denoted by $\bm{\tau}$ and $\bm{n}$, respectively. (b) The undeformed blade has dimensions $l \times b \times d$. (c) 3D sketch of the stacked blades arranged laterally along the strut in the undeformed state. The example shows five aggregates mounted on the strut, each containing $N$ blades of identical dimensions $l\times b \times d$.}
    \label{fig:blade_sketch}
\end{figure}

We consider a flexible blade with constant length $l$, width $b$, and thickness $d$, clamped at one end and initially oriented perpendicular to a uniform, steady flow $U$ of fluid with density $\rho$, as shown in \cref{fig:blade_sketch}.
It represents a flat, farmed seaweed blade from exposed areas~\citep{Buck2005}, which is the focus of this work.
The blade is assumed to be thin, i.e., $b \gg d$, and its motion is confined to the $xz$-plane.
We also assume that the structure, characterized by density $\rho_s$ and bending stiffness $EI$, is slender ($l \gg b$), and the two-dimensional inextensible Euler-Bernoulli beam model adequately represents the effects of bending.

As shown in \cref{fig:blade_sketch}a, a curvilinear coordinate $s$ is defined to denote the arc length along the blade from the base.
The position vector $\bm{r}=x(s,t)\bm{e_x}+z(s,t)\bm{e_z}$ is a function of both $s$ and time $t$, where $\bm{e_x}$ and $\bm{e_z}$ are unit vectors in the positive $x$- and $z$-directions, respectively.
The local angle $\theta$ is defined as the angle between the tangent vector $\bm{\tau}=\partial \bm{r}/\partial s$ and $-\bm{e_z}$.
Additionally, $\bm{n}$ is the normal vector.
In the following, we loosely follow \citet{Leclercq2018a} to derive the governing equation of motion for a flexible blade, with the exception that the blade is not assumed to be neutrally buoyant.

The force balance on an infinitesimal segment leads to
\begin{equation}\label{eq:force_balance}
    \mu\frac{\partial^2 \bm{r}}{\partial t^2} = \frac{\partial \bm{F}}{\partial s} + \bm{q} + \bm{B},
\end{equation}
where $\mu = \rho_s b d$ is the body mass per unit length, $\bm{q}$ is the external distributed hydrodynamic load on the body, and $\bm{B}= -(\rho_s - \rho)g b d \bm{e_z}$ is the effective weight in fluid per unit length. The internal force $F$ is expressed as $\bm{F}=T\bm{\tau} + Q\bm{n}$, where $T$ and $Q$ are the tension and shear force, respectively.
Additionally, $g$ is the acceleration of gravity.
The bending moment $M$ is related to the curvature $\kappa=\partial \theta/\partial s$ by $M=EI\kappa$.
The shear force $Q$ is given by $Q=-\partial M/\partial s=-EI\partial \kappa/\partial s$.
The boundary conditions are $x=z=\theta=0$ at the clamped base $s=0$ and $T=M=Q=0$ at the free tip $s=l$.

Let $\bm{U}=U\bm{e_x}$ be the flow velocity vector.
The relative velocity between the ambient flow and the structure can be decomposed along the blade tangential vector $\bm{\tau}$ and the blade normal vector $\bm{n}$ as
\begin{equation}
    U_{\tau} = \left(\frac{\partial \bm{r}}{\partial t} - \bm{U}\right)\cdot \bm{\tau}, \quad U_{n} = \left(\frac{\partial \bm{r}}{\partial t} - \bm{U}\right)\cdot \bm{n}.
\end{equation}

Following \citet{Leclercq2018} and \citet{Leclercq2018a}, the distributed hydrodynamic load $\bm{q}$ on the body consists of two parts, the resistive drag~\citep{Taylor1952}
\begin{equation}\label{eq:drag}
    \bm{q_{d}} = -\frac{1}{2}\rho C_D b |U_n|U_n\bm{n},
\end{equation}
and the reactive force~\citep{Lighthill1971,Candelier2011,Singh2012}
\begin{equation}\label{eq:added_mass}
    \bm{q_{am}} = -\left[\frac{\partial (m_aU_n\bm{n})}{\partial t} - \frac{\partial (m_aU_{n}U_{\tau}\bm{n})}{\partial s}  + \frac{1}{2}\frac{\partial (m_aU_n^2\bm{\tau})}{\partial s} \right].
\end{equation}
Here $m_a$ is the added mass, which may vary with $s$. In our case, we let $m_a=\pi\rho C_M b^2 / 4$ due to the constant cross-sectional profile, allowing it to be taken outside the derivatives.
This means that we also disregard viscous effects on $m_a$, which is reasonable for low to moderate $\mathrm{KC}$ numbers, where $\mathrm{KC}$ is based on the blade width $b$.
$C_D$ and $C_M$ are the local cross-sectional drag and added mass coefficients, respectively.
In principle, the values of $C_D$ and $C_M$ depend on $s$ and other flow parameters, such as the local Reynolds number or the Keulegan–Carpenter number.
Cross-sectional $C_D$ and $C_M$ are taken as constants along the blade in this work.
In \citet{Leclercq2018} and \citet{Leclercq2018a}, $C_M = 1$ is universally assumed.
In \cref{sec:applicability_hydro_load}, we justify the applicability of the hydrodynamic force model, i.e., \cref{eq:drag} and \cref{eq:added_mass}, to side-by-side blades, as illustrated in \cref{fig:blade_sketch}c, within a confined fluid domain, such as the experimental tank.

For an inextensible structure, \cref{eq:added_mass} simplifies to
\begin{equation}\label{eq:added_mass_inextesible}
    \bm{q_{am}} = -m_a\left[\frac{\partial^2 \bm{r}}{\partial t^2}\cdot\bm{n} - 2\frac{\partial\theta}{\partial t}U_{\tau} + \kappa\left(U^2_{\tau} - \frac{1}{2}U^2_n\right)\right]\bm{n}.
\end{equation}
For a detailed derivation, interested readers are referred to Appendix A of \citet{Leclercq2018a}.
The Froude-Krylov force is not included as this work focuses on steady ambient flows.

Using the relation between the bending moment $M$ and the shear force $Q$, \cref{eq:force_balance} becomes
\begin{equation}\label{eq:dimensional_ge}
    \mu\frac{\partial^2 \bm{r}}{\partial t^2} = \frac{\partial}{\partial s}\left(T + \frac{1}{2}EI\kappa^2\right)\bm{\tau} + \left(\kappa T - EI\frac{\partial^2 \kappa}{\partial s^2}\right)\bm{n} + \bm{q_d} + \bm{q_{am}} + \bm{B}.
\end{equation}
Projection of \cref{eq:dimensional_ge} on the tangential vector yields
\begin{equation}\label{eq:dimensional_ge_tangential}
    \mu\frac{\partial^2 \bm{r}}{\partial t^2}\cdot \bm{\tau} = \frac{\partial}{\partial s}\left(T + \frac{1}{2}EI\kappa^2\right) + \bm{B}\cdot \bm{\tau} = \frac{\partial}{\partial s}\left(T + \frac{1}{2}EI\kappa^2\right) + (\rho_s - \rho)g b d \cos\theta.
\end{equation}
Integrating \cref{eq:dimensional_ge_tangential} over $l$ to $s$ and combining the boundary conditions $T(s=l)=\kappa(s=l)=0$ provides
\begin{equation}\label{eq:tension_explicit}
    T = \mu \int^s_l \frac{\partial^2 \bm{r}}{\partial t^2} \cdot \bm{\tau}\mathrm{d}s - \frac{1}{2}EI \kappa^2 - (\rho_s - \rho)g b d \int^s_l \cos\theta\mathrm{d}s.
\end{equation}
Projection of \cref{eq:dimensional_ge} on the normal vector yields
\begin{equation}\label{eq:dimensional_ge_normal}
\begin{split}
    \mu\frac{\partial^2 \bm{r}}{\partial t^2} \cdot \bm{n} &= \kappa T - EI\frac{\partial^2 \kappa}{\partial s^2} + \left(\bm{q_d} + \bm{q_{am}} + \bm{B}\right)\cdot \bm{n} \\
    &= \kappa T - EI\frac{\partial^2 \kappa}{\partial s^2} - \frac{1}{2}\rho C_D b |U_n|U_n - m_a \frac{\partial^2 \bm{r}}{\partial t^2}\cdot\bm{n} - m_a \left[\kappa\left(U^2_{\tau} - \frac{1}{2}U^2_n\right)-2U_{\tau}\frac{\partial \theta}{\partial t}\right] - (\rho_s - \rho) g b d \sin\theta.
\end{split}
\end{equation}
Substituting the expression for tension $T$ from \cref{eq:tension_explicit} into \cref{eq:dimensional_ge_normal} and rearranging provides the governing equation of motion for a flexible blade:
\begin{equation}\label{eq:ge}
    \begin{split}
        &(\mu + m_a) \frac{\partial^2 \bm{r}}{\partial t^2} \cdot \bm{n} - \mu\kappa \int^s_l \frac{\partial^2 \bm{r}}{\partial t^2} \cdot \bm{\tau}\mathrm{d}s + EI\left(\frac{\partial^2 \kappa}{\partial s^2} + \frac{1}{2}\kappa^3\right) \\
        &\quad + m_a\left[\kappa\left(U^2_{\tau} - \frac{1}{2}U^2_n\right)-2U_{\tau}\frac{\partial \theta}{\partial t}\right] + \frac{1}{2}\rho C_D b |U_n|U_n + (\rho_s - \rho) g b d \left(\kappa \int^s_l \cos\theta\mathrm{d}s + \sin\theta\right) = 0.
    \end{split}
\end{equation}

Following \citet{Leclercq2018a}, \cref{eq:ge} is made non-dimensional using the blade length $l$ and the first natural period of the structure in small-amplitude oscillations in the fluid $T_n=l^2\sqrt{(\mu + m_a)/EI}$.
The non-dimensional form is
\begin{equation}\label{eq:non_ge}
    \begin{split}
        &\frac{\partial^2 \tilde{\bm{r}}}{\partial \tilde{t}^2} \cdot \bm{n} - (1 - \upbeta)\tilde{\kappa} \int^{\tilde{s}}_1 \frac{\partial^2 \tilde{\bm{r}}}{\partial \tilde{t}^2} \cdot \bm{\tau}\mathrm{d}\tilde{s} + \frac{\partial^2 \tilde{\kappa}}{\partial \tilde{s}^2} + \frac{1}{2}\tilde{\kappa}^3 \\
        &\quad + \upbeta\left[\tilde{\kappa}\left(\widetilde{U}^2_{\tau} - \frac{1}{2}\widetilde{U}^2_n\right)-2\widetilde{U}_{\tau}\frac{\partial \theta}{\partial \tilde{t}}\right] + \upbeta \uplambda |\widetilde{U}_n|\widetilde{U}_n + \mathrm{B}\left(\tilde{\kappa}\int^{\tilde{s}}_1 \cos{\theta}\mathrm{d}\tilde{s} + \sin{\theta}\right) = 0,
    \end{split}
\end{equation}
where a tilde $(\,\tilde{}\,)$ denotes non-dimensional variables.
The boundary conditions are $\tilde{\bm{r}}=\bm{0}$ and $\theta = 0$ at $\tilde{s}=0$ and $\tilde{\kappa} = \partial\tilde{\kappa}/\partial\tilde{s} = 0$ at $\tilde{s} = 1$.
The non-dimensional relative velocity is given by $\widetilde{U}_n\bm{n} + \widetilde{U}_{\tau}\bm{\tau} = \partial\tilde{\bm{r}}/\partial\tilde{t} - \mathrm{u}/\sqrt{\upbeta}\bm{e_x}$, where $\mathrm{u}=\sqrt{\mathrm{Ca}/\uplambda}$ is the reduced velocity.

The system is primarily governed by four non-dimensional parameters:
\begin{equation}\label{eq:non_dimensional_parameters}
    \upbeta = \frac{m_a}{\mu + m_a},\quad \mathrm{Ca} = \frac{1}{2}\frac{\rho C_D b U^2 l^3}{EI},\quad \mathrm{B} = \frac{(\rho_s - \rho)g b d l^3}{EI},\quad \uplambda = \frac{\rho C_D b l}{2 m_a} = \frac{2}{\pi}\frac{C_D}{C_M} \frac{l}{b}.
\end{equation}
Physically, the mass ratio $\upbeta$ represents the proportion of fluid inertia relative to the total inertia of the system. When the blades are very thin, i.e., $d \ll b$, $\upbeta$ approaches unit $(\upbeta \to 1)$.
The Cauchy number $\mathrm{Ca}$ represents the ratio of the drag force to the restoring force due to bending stiffness.
Although $\mathrm{Ca}$ does not explicitly appear in \cref{eq:non_ge}, it is incorporated through the reduced velocity $\mathrm{u}=\sqrt{\mathrm{Ca}/\uplambda}$.
The buoyancy parameter $\mathrm{B}$ indicates the ratio of the restoring force due to buoyancy to that due to bending stiffness. For neutrally buoyant structures, i.e. $\mathrm{B} = 0$, \cref{eq:non_ge} reduces to the same form presented by \citet{Leclercq2018a}.
We refer to $\mathrm{B}$ as the buoyancy parameter following \citet{Lei2019a} and \citet{Vettori2024}; however, in this study, the structure's density is greater than that of water.
The slenderness parameter $\uplambda$ is proportional to the aspect ratio $l/b$ and can be interpreted as the ratio of the resistive drag to the reactive force.
Here we choose $\uplambda$ instead of $l/b$ as the non-dimensional parameters, since the hydrodynamic coefficients $C_D$ and $C_M$ are naturally incorporated into $\uplambda$ from the non-dimensionalization of the governing equations~\cref{eq:non_ge}.
Similar approaches have been adopted in previous studies~\citep{Luhar2011,Leclercq2018a,Leclercq2018,Zhang2020,Lei2021,Schaefer2024}, where $C_D$ or $C_M$ are embedded in the non-dimensional parameters.

In the experiments, blades were assembled into aggregates containing one or more layers; the aggregates were then spaced laterally along the strut, as shown in \cref{fig:blade_sketch}c. Each aggregate comprised $N$ blades of identical dimensions $l\times b \times d$.
The blades were neatly overlapped on one another and, although not physically adhered, they behaved similarly.
As a first approximation, when studying an aggregate of blades, we treat the blades as a single unit and apply the same theory from \cref{sec:analytical_model}, but with the blade thickness $d$ replaced by $Nd$ and the bending stiffness $EI$ replaced by $NEI$, where $N$ is the number of overlapping blades in one aggregate.
Consider $N$ blades stacked neatly in one aggregate and clamped at one end.
Assuming negligible inter-blade shear transfer so that all blades share the same curvature $\kappa$ along the blades, the total bending moment is $N EI \kappa$.
Hence the aggregate’s equivalent bending stiffness is $N EI$.
The quantities $Nd$ and $N EI$ are referred to as the equivalent thickness and equivalent bending stiffness of the aggregate of blades.
According to \cref{eq:non_dimensional_parameters}, given the same values of $C_D$ and $C_M$, for an aggregate consisting of $N$ blades, the buoyancy parameter $\mathrm{B}$ and the slenderness parameter $\uplambda$ are identical to those of an individual blade within that aggregate.
The Cauchy number $\mathrm{Ca}$ is reduced to $1/N$ of that for an individual blade due to the increased bending stiffness $EI$, and the mass ratio $\upbeta$ can be rewritten as
\begin{equation}\label{eq:mass_ratio_N_layers}
    \upbeta = \frac{m_a}{\mu + m_a} = \frac{1}{1 + \dfrac{\mu}{m_a}} = \frac{1}{1 + \dfrac{N \rho_s b d}{\pi \rho C_M b^2/4}} = \frac{1}{1 + \dfrac{1}{C_M}\dfrac{4 N}{\pi}\dfrac{\rho_s}{\rho}\dfrac{d}{b}},
\end{equation}
which decreases slightly as $\mu$ increases to $N\mu$ and $m_a$ remains unchanged (assuming the same $C_M$).

\subsection{Drag reduction in the static regime}\label{sec:drag_reduction}
A flexible blade may experience less drag than a rigid blade of the same geometry, a phenomenon known as drag reduction.
In the theoretical analysis, the reconfiguration number $\mathcal{R}$ and the effective length ratio $l_e/l$ are used to quantify the extent of drag.
In the static regime prior to flutter, an analytical prediction of drag reduction can be made by removing the unsteady terms from \cref{eq:non_ge} and solving the resulting boundary-value problem.

\subsubsection{Reconfiguration number and effective length}
Drag reduction can be quantified by using the reconfiguration number~\citep{Gosselin2010} or the effective length~\citep{Luhar2011}.
The reconfiguration number $\mathcal{R}$ is defined as the ratio of the drag force on the deflected structure to that on a rigid counterpart of the same initial geometry.
The effective blade length $l_e$ is defined as the length of a rigid, vertical blade that would generate the same horizontal drag as the flexible blade of total length $l$.
These two definitions are essentially equivalent, as expressed by:
\begin{equation}\label{eq:reconfiguration_number_effective_length}
    \mathcal{R} = \frac{l_e}{l} = \frac{F}{F_{\mathrm{rigid}}} = \frac{F}{0.5\rho C_D b l U^2},
\end{equation}
where $U$ is the ambient horizontal velocity, $F$ is the total horizontal drag on the deflected blade, and $F_{\mathrm{rigid}}=0.5\rho C_D b l U^2$ is the total horizontal drag on the rigid, upright blade of the same dimensions.
There are multiple ways to express $F$, consequently, leading to different approaches for calculating $\mathcal{R}$ or $l_e/l$ in the literature.

In the static regime, following \citet{Luhar2011}, who applied the cross-flow principle and strip theory, the total drag force can be expressed as
\begin{equation}\label{eq:drag_static_regime}
    F = \int^l_0 \frac{1}{2} \rho C_D b (U \cos\theta)^2 \cos\theta \, \mathrm{d}s,
\end{equation}
where $U\cos\theta$ is the local normal velocity accounting for the streamlined shape of the blade, and the final $\cos\theta$ term in the integrand accounts for the horizontal component of the drag force or the reduced frontal area.
In dimensionless form, using $\tilde{s}=s/l$ as in \cref{sec:analytical_model}, the effective length is given by
\begin{equation}\label{eq:effective_length}
    \frac{l_e}{l} = \int^1_0 \cos^3\theta \, \mathrm{d}\tilde{s},
\end{equation}
which can also be used to calculate $\mathcal{R}$ in the static regime, as in \citet{Gosselin2010}.
We denote this definition as $\mathcal{R}^\mathrm{ana,g}$, where the superscript~$^\mathrm{g}$ stands for the \textit{global} integral form over the interval $0 \leq \tilde{s} \leq 1$.
The concept of effective length has been more extensively used in the context of oscillatory flows in later studies~\citep{Luhar2016,Luhar2017,Lei2019}.

In \cref{eq:reconfiguration_number_effective_length}, $F$ is also equal to the internal shear force at the base, i.e.,
\begin{equation}\label{eq:external_equal_shear_at_the_base}
    F = Q(s=0) = -EI\left.\frac{\partial \kappa}{\partial s}\right|_{s=0}.
\end{equation}
Inserting \cref{eq:external_equal_shear_at_the_base} into \cref{eq:reconfiguration_number_effective_length} gives:
\begin{equation}
    \mathcal{R} = - \left.\frac{1}{\mathrm{Ca}}\frac{\partial\tilde{\kappa}}{\partial \tilde{s}}\right|_{s=0},
\end{equation}
as detailed in \citet{Leclercq2018a}.
This definition is denoted as $\mathcal{R}^\mathrm{ana,l}$, where the superscript~$^\mathrm{l}$ represents the \textit{local} horizontal force balance at $\tilde{s}=0$.

In the post-critical regime, where flutter occurs, \cref{eq:drag_static_regime,eq:effective_length} are no longer applicable due to the blade's dynamic motion contributing to the horizontal force.
In the experiments, it is more practical to use the measured horizontal force as $F$ in \cref{eq:reconfiguration_number_effective_length}, and the corresponding reconfiguration number is denoted as $\mathcal{R}^{\mathrm{exp.}}$.
Theoretically, $F$ is equal to the horizontal component of the total external force.
For the remainder of this study, we will use $\mathcal{R}$ instead of $l_e/l$ to represent drag reduction without loss of generality.

\subsubsection{Static governing equation prior to flutter}\label{sec:drag_reduction_static}
For velocities lower than the critical value at which flutter starts to occur, the blade will reach a static equilibrium position.
Removing all time derivatives from \cref{eq:non_ge} gives
\begin{equation}\label{eq:static_non_ge_00}
    \frac{\partial^2 \tilde{\kappa}}{\partial \tilde{s}^2} + \frac{1}{2}\tilde{\kappa}^3 + \upbeta \tilde{\kappa}\left(\widetilde{U}^2_{\tau} - \frac{1}{2}\widetilde{U}^2_n\right) + \upbeta \uplambda |\widetilde{U}_n|\widetilde{U}_n + \mathrm{B}\left(\tilde{\kappa}\int^{\tilde{s}}_1 \cos{\theta}\mathrm{d}\tilde{s} + \sin{\theta}\right) = 0.
\end{equation}
In the static regime, we have
\begin{equation}
    \widetilde{U}_n\bm{n} + \widetilde{U}_{\tau}\bm{\tau} = -\frac{\mathrm{u}}{\sqrt{\upbeta}}\bm{e_x} = -\sqrt{\frac{\mathrm{Ca}}{\upbeta\uplambda}}\bm{e_x},
\end{equation}
which leads to
\begin{equation}\label{eq:relative_velocities}
    \widetilde{U}_n = -\sqrt{\frac{\mathrm{Ca}}{\upbeta\uplambda}}\bm{e_x}\cdot\bm{n} = -\sqrt{\frac{\mathrm{Ca}}{\upbeta\uplambda}}\cos\theta,\quad \widetilde{U}_{\tau} = -\sqrt{\frac{\mathrm{Ca}}{\upbeta\uplambda}}\bm{e_x}\cdot\bm{\tau} = -\sqrt{\frac{\mathrm{Ca}}{\upbeta\uplambda}}\sin\theta.
\end{equation}

Inserting the expressions above and $\tilde{\kappa}=\partial \theta/\partial \tilde{s}$ into \cref{eq:static_non_ge_00} yields the static equation
\begin{equation}\label{eq:non_static_ge}
    \underbrace{\frac{\partial^3 \theta}{\partial \tilde{s}^3} + \frac{1}{2}\left(\frac{\partial \theta}{\partial \tilde{s}}\right)^3}_\text{bending stiffness} + \underbrace{\frac{\mathrm{Ca}}{\uplambda} \frac{\partial \theta}{\partial \tilde{s}}\left(\sin^2\theta - \frac{1}{2}\cos^2\theta\right)}_\text{reactive force} - \underbrace{\bigg(\mathrm{Ca}\cos^2{\theta}\bigg)}_\text{drag}+ \underbrace{\mathrm{B}\left(\frac{\partial \theta}{\partial\tilde{s}}\int^{\tilde{s}}_1 \cos\theta\mathrm{d}\tilde{s} + \sin{\theta}\right)}_\text{buoyancy} = 0,
\end{equation}
with boundary conditions $\theta=0$ at $\tilde{s}=0$ and $\partial\theta/\partial \tilde{s} = \partial^2\theta/\partial \tilde{s}^2 = 0$ at $\tilde{s}=1$.

\citet{Luhar2011} omitted the second bending stiffness term and the reactive force terms in \cref{eq:non_static_ge}, while \citet{Gosselin2010} further assumed $\mathrm{B}=0$, leading to an inconsistency between the definitions of $\mathcal{R}^{\mathrm{ana,g}}$ and $\mathcal{R}^{\mathrm{ana,l}}$, i.e.,
\begin{equation}
     \int^1_0 \cos^3\theta \, \mathrm{d}\tilde{s} = \mathcal{R}^{\mathrm{ana,g}} \neq \mathcal{R}^{\mathrm{ana,l}} = -\left.\frac{\partial\tilde{\kappa}}{\partial \tilde{s}}\right|_{\tilde{s}=0}.
\end{equation}
Therefore, caution is needed when comparing $\mathcal{R}^{\mathrm{ana,g}}$ derived using solutions from \citet{Luhar2010} and \citet{Gosselin2010} with experimental data.
In the experiments, the forces at the base of the blade instead of the total forces on the blade are measured.
Additionally, tension effects are always present in reality.
Consequently, when comparing the analytical prediction with experimental results, $\mathcal{R}^{\mathrm{ana,l}}$ would be a more appropriate choice than $\mathcal{R}^{\mathrm{ana,g}}$.

To solve \cref{eq:non_static_ge}, which is a boundary-value problem, we follow the approach of \citet{Leclercq2018a} and \citet{Leclercq2018}.
We first discretize the blade using the Gauss-Lobatto distribution $s_k=1/2[1- \cos((k-1)/(N-1)\pi)]$ along with the boundary conditions.
The derivatives and integrals are then computed using Chebyshev collocation and the Clenshaw-Curtis quadrature formulas, respectively. Even though the solution to \cref{eq:non_static_ge} in this subsection also involves numerical integrations, we will still refer to it as an analytical solution to distinguish it from the time-domain numerical solutions provided by the truss-spring solver described later in \cref{sec:numerical_model}.

\subsection{Reactive force model and stability}
Flutter is anticipated at large $\mathrm{Ca}$ numbers.
Determining the exact stability threshold requires a full stability analysis of the nonlinear equation of motion \cref{eq:non_ge}.
Numerical efforts by \citet{Leclercq2018a} addressed this.
Here, we introduce qualitative approaches to identify which terms in the reactive force model precipitate flutter.

By tracing the origin of the terms in \cref{eq:non_ge}, we identify how each term in \cref{eq:added_mass} contributes to \cref{eq:non_ge} and to the static terms in \cref{eq:non_static_ge}.
We assume constant $C_M$ and take it outside of the derivatives in \cref{eq:added_mass}.
For $\bm{q}_{\bm{am},1}$, we have
\begin{equation}
    \bm{q}_{\bm{am},1} = \underbrace{-m_a \frac{\partial (U_n\bm{n})}{\partial t}}_{\mathrm{In~\cref{eq:added_mass}}} = \underbrace{-m_a \left[\frac{\partial^2\bm{r}}{\partial t^2}\cdot\bm{n} - \frac{\partial\theta}{\partial t}U_{\tau}\right]}_{\mathrm{In~\cref{eq:added_mass_inextesible}}} \sim \underbrace{\frac{\partial^2 \tilde{\bm{r}}}{\partial \tilde{t}^2} \cdot \bm{n} - \upbeta \widetilde{U}_{\tau}\frac{\partial \theta}{\partial \tilde{t}}}_{\mathrm{In~\cref{eq:non_ge}}}.
\end{equation}
In the static limit, $\bm{q}_{\bm{am},1}$ vanishes and contributes nothing to the static balance.

For $\bm{q}_{\bm{am},2}$, we have
\begin{equation}
    \bm{q}_{\bm{am},2} = \underbrace{m_a \frac{\partial (U_{\tau}U_{n}\bm{n})}{\partial s}}_{\mathrm{In~\cref{eq:added_mass}}} = \underbrace{-m_a \left[- \frac{\partial\theta}{\partial t}U_{\tau} + \kappa U^2_{\tau}\right]}_{\mathrm{In~\cref{eq:added_mass_inextesible}}} \sim \underbrace{\upbeta\left[-\widetilde{U}_{\tau}\frac{\partial \theta}{\partial \tilde{t}} + \tilde{\kappa} \widetilde{U}^2_{\tau}\right]}_{\mathrm{In~\cref{eq:non_ge}}}.
\end{equation}
Removing the unsteady terms and substituting $\widetilde{U}_{\tau}$ using \cref{eq:relative_velocities} gives the contribution to the static balance:
\begin{equation}
    \bm{q}_{\bm{am},2} \to \frac{\mathrm{Ca}}{\uplambda}\frac{\partial\theta}{\partial\tilde{s}}\sin^2\theta \text{ in \cref{eq:non_static_ge}}.
\end{equation}

For $\bm{q}_{\bm{am},3}$, we have
\begin{equation}
    \bm{q}_{\bm{am},3} = \underbrace{-\frac{1}{2}m_a\frac{\partial (U_n^2\bm{\tau})}{\partial s}}_{\mathrm{In~\cref{eq:added_mass}}} = \underbrace{-m_a \left[-\frac{1}{2}\kappa U^2_{n}\right]}_{\mathrm{In~\cref{eq:added_mass_inextesible}}} \sim \underbrace{\upbeta\left[-\frac{1}{2}\widetilde{\kappa}\widetilde{U}^2_{n}\right]}_{\mathrm{In~\cref{eq:non_ge}}}.
\end{equation}
Similarly, substituting $\widetilde{U}_{n}$ using \cref{eq:relative_velocities} gives the contribution to the static balance:
\begin{equation}
    \bm{q}_{\bm{am},3} \to -\frac{1}{2}\frac{\mathrm{Ca}}{\uplambda}\frac{\partial\theta}{\partial\tilde{s}}\cos^2\theta  \text{ in \cref{eq:non_static_ge}}.
\end{equation}

There are two terms in $\bm{q}_{\bm{am},2}$, which are further denoted by
\begin{equation}
    \bm{q}_{\bm{am},2,1} = -\upbeta \widetilde{U}_{\tau}\frac{\partial \theta}{\partial \tilde{t}},\quad \bm{q}_{\bm{am},2,2} = \upbeta\tilde{\kappa} \widetilde{U}^2_{\tau}.
\end{equation}

We choose to retain the first term $\bm{q}_{\bm{am},1}$ as the well-known added mass term.
Making use of the numerical model later introduced in \cref{sec:numerical_model} and testing different combinations of including or excluding the terms $\bm{q}_{\bm{am},2,1}$, $\bm{q}_{\bm{am},2,2}$, and $\bm{q}_{\bm{am},3}$ in the hydrodynamic load model, the corresponding numerical reconfiguration number $\mathcal{R}$ and critical $\mathrm{Ca}$ can be obtained.

Another qualitative approach to analyse the stability of the system is to map the reactive force model to the well-known linear small-amplitude flutter equation for an undamped beam in axial flow.
We first introduce the reduced velocity:
\begin{equation}
    \mathrm{u} = \sqrt{\frac{\mathrm{Ca}}{\uplambda}} = Ul\sqrt{\frac{m_a}{EI}},
\end{equation}
which is commonly used when discussing features of flutter instability.
The reduced velocity represents the relative magnitude of the reactive fluid force and the structural stiffness determines the onset of flutter, as noted by \citet{Leclercq2018a}.

The classic non-dimensional small-amplitude flutter equation for an undamped beam in axial flow~\citep{Paidoussis1998} reads
\begin{equation}\label{eq:linear_beam_flutter}
    \frac{\partial^2 \tilde{\eta}}{\partial \tilde{t}^2} + 2 \mathrm{u} \sqrt{\upbeta} \frac{\partial^2 \tilde{\eta}}{\partial \tilde{t} \partial \tilde{s}} + \mathrm{u}^2 \frac{\partial^2\tilde{\eta}}{\partial \tilde{s}^2} + \frac{\partial^4\tilde{\eta}}{\partial \tilde{s}^4} = 0,
\end{equation}
with boundary conditions $\tilde{\eta}=\partial\tilde{\eta}/\partial\tilde{s}= 0$ at $\tilde{s}=0$ and $\partial^2\tilde{\eta}/\partial\tilde{s}^2=\partial^3\tilde{\eta}/\partial\tilde{s}^3=0$ at $\tilde{s}=1$,
where $\tilde{\eta}(\tilde{s},\tilde{t})$ is the normalized lateral displacement, and the other symbols have the same definition as in this paper.
By comparing the terms in the reactive force with the terms in \cref{eq:linear_beam_flutter}, we find the following relations:
\begingroup
\allowdisplaybreaks
\begin{alignat*}{3}
    &\bm{q}_{\bm{am},1} && \sim \frac{\partial^2 \tilde{\bm{r}}}{\partial \tilde{t}^2} \cdot \bm{n} - \upbeta \widetilde{U}_{\tau}\frac{\partial \theta}{\partial \tilde{t}} && \sim \frac{\partial^2 \tilde{\eta}}{\partial \tilde{t}^2} + \mathrm{u} \sqrt{\upbeta} \frac{\partial }{\partial t}\left(\frac{\partial \tilde{\eta}}{\partial \tilde{s}}\right), \\
    &\bm{q}_{\bm{am},2,1} &&\sim -\upbeta \widetilde{U}_{\tau}\frac{\partial \theta}{\partial \tilde{t}} && \sim \mathrm{u} \sqrt{\upbeta} \frac{\partial^2 \tilde{\eta}}{\partial \tilde{t} \partial \tilde{s}} = \mathrm{u} \sqrt{\upbeta} \frac{\partial }{\partial t}\left(\frac{\partial \tilde{\eta}}{\partial \tilde{s}}\right), \\
    &\bm{q}_{\bm{am},2,2} && \sim \upbeta\tilde{\kappa} \widetilde{U}^2_{\tau} = \upbeta\widetilde{U}^2_{\tau}\frac{\partial\theta}{\partial\tilde{s}} && \sim \mathrm{u}^2 \frac{\partial^2\tilde{\eta}}{\partial \tilde{s}^2} = \mathrm{u}^2 \frac{\partial}{\partial s}\left(\frac{\partial\tilde{\eta}}{\partial \tilde{s}}\right),\\
    &\bm{q}_{\bm{am},3} &&\sim -\frac{1}{2}\upbeta\tilde{\kappa}\widetilde{U}^2_{n} &&\sim 0.
\end{alignat*}
\endgroup
According to the small-amplitude motion assumption, $\partial\tilde{\eta}/\partial\tilde{s}\simeq\tan\alpha\simeq\alpha$, where $\alpha=\pi/2-\theta$.
The relation can also be found in \citet{Leclercq2018a}.
Their appendix demonstrates that \cref{eq:linear_beam_flutter} cannot be used to predict the onset of flutter.
This limitation arises from the presence of drag, which is not included in \cref{eq:linear_beam_flutter}. \citet{Leclercq2018a} noted that, at the static equilibrium position, the drag perturbation scales with $\uplambda\cos\theta$, which remains of order $O(1)$ as $\uplambda\to\infty$ and $\theta\to \pi/2$. As a result, the contribution of the resistive drag remains significant even on the part of the structure that is nearly parallel to the flow.
Since the stability theory of \cref{eq:linear_beam_flutter} has been well developed~\citep{Paidoussis1998}, we can still use it to qualitatively examine the different terms in the reactive force model concerning system stability.

\section{Experiments}\label{sec:expriments}
\begin{figure}
    \centerline{\includegraphics{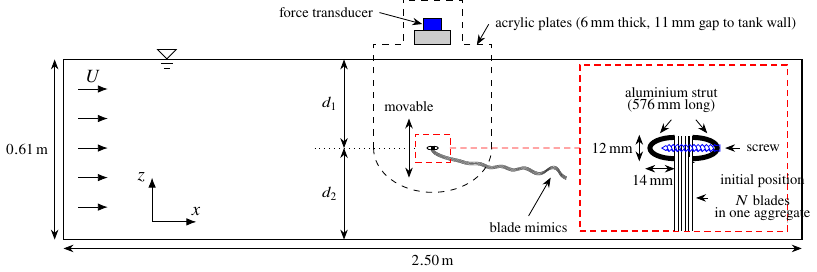}}
    \caption{Side view sketch of the test section and setup in the Circulating Water Tunnel (CWT) at NTNU. An enlarged view of the strut and the stacked blades in their initial positions is also provided.}
    \label{fig:exp_sketch}
\end{figure}

\begin{figure}
    \centerline{\includegraphics[width=1.0\textwidth]{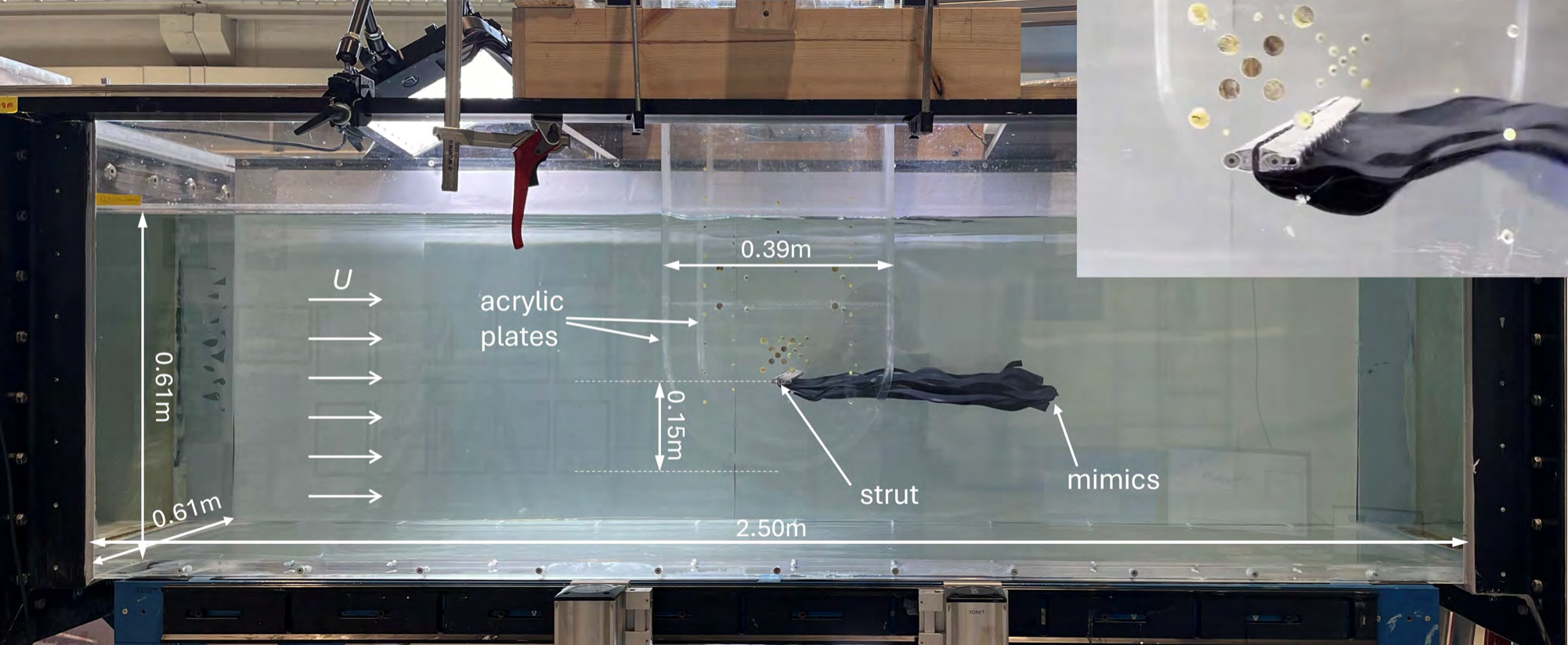}}
    \caption{Photo of the test section of the Circulating Water Tunnel (CWT) at NTNU and the experimental setup. A close view of the strut is shown in the upper right corner.}
    \label{fig:exp_photo}
\end{figure}

The experiments were performed in the Circulating Water Tunnel (CWT) at the Norwegian University of Science and Technology (NTNU) in Trondheim.
An illustration of the test setup is provided in \cref{fig:exp_sketch} and \cref{fig:exp_photo}.
The test section of the CWT is \SI{2.50}{m} long, \SI{0.61}{m} wide, and \SI{0.61}{m} deep.

The blade mimics were clamped at the top end in an aluminum strut positioned in the middle of the tank.
Blade mimics were fixed between the strut using twenty-three screws distributed along the strut, as illustrated in \cref{fig:exp_sketch} and \cref{fig:exp_photo}.
The strut was cut from a \SI{576}{mm} long cylinder with an elliptical cross-section, featuring a semi-major axis of \SI{14}{mm} and a semi-minor axis of \SI{6}{mm}.
The strut was secured to two side acrylic plates, each with a thickness of \SI{6}{mm}.
A \SI{13}{mm}-wide streamlined fairing was applied along the plate perimeter to reduce drag and flow disturbances.
A rig connected the two plates at the top, where a six-degree-of-freedom (6-DOF) force transducer was mounted.
An HBM MCS10 multicomponent sensor (BG2 size, type 010) was used. The measurement ranges were \SI{2}{kN} for the lateral forces $F_x$ and $F_y$, \SI{10}{kN} for the axial force $F_z$, and \SI{0.15}{kN.m} for the bending moments $M_x$, $M_y$, and the torsional moment $M_z$.
With an accuracy class of 0.2, the sensor provides an overall measurement uncertainty on the order of \SI{4}{N} in lateral forces, \SI{20}{N} in axial force, and \SI{0.30}{N.m} in moments.
Given a lever arm $L=\SI{0.65}{m}$, the moment measurements have an accuracy of $\SI{0.46}{N}$ in the streamline force, which is better than the force measurements.
We base our analysis of the bulk drag coefficient on $M_y$ instead of $F_x$ for a better accuracy.
The sensor was calibrated by the laboratory staff before the experiments.
A \SI{30}{s} zero balance was applied before each measurement after the tank was quiescent.
Gravity of the blade mimics and the supporting frames was not recorded and any static contribution from the frame or blade weight to the horizontal force was removed.
All six components were sampled at \SI{200}{Hz}.
Before each measurement, we verified that the lateral force (along the strut) was approximately zero at the maximum velocity $U=\SI{0.975}{m/s}$ to avoid misalignment between the force transducer and the experimental facility.

We used flat blade mimics to represent \textit{Laminaria saccharina} blades cultivated in exposed areas, which are reported to be flat and narrow~\citep{Buck2005}.
Seaweed morphology varies a lot in nature~\citep{Neufeldt2025}.
Sets of four different blade mimics with rectangular cross-sections were fabricated from silicone rubber sheets, with a density of $\rho_s = \SI[separate-uncertainty=true]{1264.5(66.0)}{kg/m^3}$ and Young's modulus of $E = \SI[separate-uncertainty=true]{5.36(40)}{MPa}$.
The density was determined by measuring the mass and volume of seven rectangular samples. Mass measurements were conducted using a digital scale with a precision of \SI{0.01}{g}, while volume was calculated based on the dimensions obtained from a steel ruler accurate to \SI{1}{mm} for length and width, and a digimatic Mitutoyo caliper for thickness.
Young's modulus was measured using Peirce's testing apparatus~\citep{Peirce1930}.
The dimensions of the four mimics are provided in \cref{tab:mimic_dimensions}.
The material properties are consistent with the measured values  of cultivated seaweeds in \citet{Fredriksson2020} and \citet{Zhu2021}.
All four mimic types had the same length. Mimics 1, 2, and 3 shared the same thickness $d$ but differed in width $b$, while Mimics 2 and 4 had the same width $b$ but differed in thickness $d$.

Along the strut, several blade aggregates were placed side by side, as shown in \cref{tab:mimic_dimensions} and \cref{fig:blade_sketch}c.
Within each aggregate, a number of blades ($N=10,5,2,1$) with the same shape were neatly overlapped on top of one another (see the magnified window in \cref{fig:exp_sketch}).
This arrangement allowed us to study the effect of line density, i.e., the number of blades per unit length.

\begin{table}
\begin{center}
    \caption{Dimensions of the four blade mimics. The final column gives the number of aggregates mounted along the strut.}
    \begin{tabular}{lccccc}
        \toprule
        Mimic & Length $l$ (\si{m}) & Width $b$ (\si{mm}) & Thickness  $d$ (\si{mm}) & Aspect ratio $l/b$  & Number of aggregates ($b_\mathrm{strut}/b$) \\
        \midrule
        1     & 0.5          & ~26.2         & 0.5               & 19.1      &   22    \\
        2     & 0.5          & ~52.4         & 0.5               & ~9.5      &   11   \\
        3     & 0.5          & 115.2         & 0.5               & ~4.3      &   ~5  \\
        4     & 0.5          & ~52.4         & 1.0               & ~9.5      &   11   \\
        \bottomrule
    \end{tabular}
    \label{tab:mimic_dimensions}
\end{center}
\end{table}

\begin{figure}
    \centerline{\includegraphics{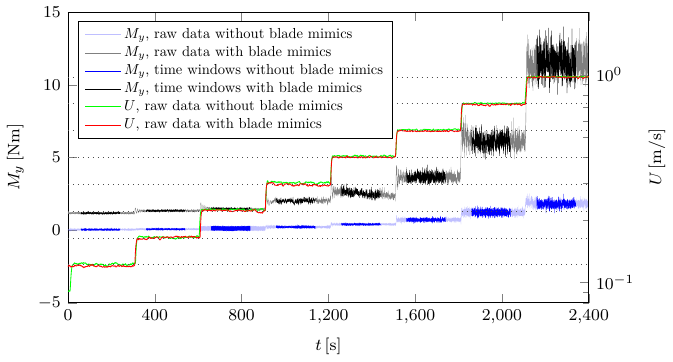}}
    \caption{Time-series examples of current speeds and measured moments for the case with Mimic 3 and $N=10$ in one aggregate and one base case without blade mimics at $d_1=\SI{18}{cm}$. The horizontal dotted lines represent the target current speeds. The forces are plotted in linear scale on the left axis and the flow speed is plotted in logarithmic scale on the right axis. The corresponding mean and standard deviation are provided in \cref{tab:mean_and_sd}.}
    \label{fig:raw_data_and_time_windows}
\end{figure}

A total of eleven current speeds, $U=0.050$, $0.067$, $0.091$, $0.122$, $0.164$, $0.221$, $0.297$, $0.400$, $0.538$, $0.725$, and $\SI{0.975}{m/s}$ were studied.
The speeds were spaced evenly on a logarithmic scale.
The actual current speed $U$ might deviate slightly from the target speeds mentioned above, as shown in \cref{fig:raw_data_and_time_windows}.
The flow speeds used for analysis were calibrated based on the pump rotation frequency, and this calibration was performed prior to the tests.
The CWT is equipped with a honeycomb flow straightener and a contraction before the test section to suppress turbulence intensity to around \SI{1}{\percent}, i.e., velocity fluctuations of roughly \SI{1}{\percent} of the free-stream velocity in the empty tunnel.
The frame (including the strut and the acrylic plates) holding the blades was vertically adjustable.
For each combination of flow speed and blade mimic, the submergence $d_1$ of the strut and the distance $d_2$ to the tank bottom were adjusted to ensure that the blade tip would not touch the bottom and the blades would remain as far from the free surface as possible, to minimize free-surface effects.
The possible submergence depths were $d_1=30,\, 26,\, 22,\, 18,\, 14$ and $\SI{10}{cm}$.
In some cases, tests were performed at multiple submergence depths.
A complete list is given in \cref{tab:submergence_depth}. Free-surface disturbances became noticeable at small submergence depths, but overall the submergence had only a minor influence on the results.
The water temperature was \SIrange{23.0}{25.5}{\celsius}.

\begin{table}
    \centering
    \caption{Submergence depth ($d_1$) of the strut.}
    \footnotesize
    \begin{minipage}[t]{0.32\textwidth}
        \centering
        \begin{tabular}{lrcc}
            \toprule
            Mimic & $N$ & $d_1$ (\si{cm}) & $U$ (\si{m/s}) \\
            \midrule
            1 & 10 & 14 & 0.050 $\sim$ 0.975  \\
            1 & 10 & 10 & 0.050 $\sim$ 0.975  \\
            1 & 10 & 30 & 0.221 $\sim$ 0.975  \\
            1 & 5  & 30 & 0.050 $\sim$ 0.975  \\
            1 & 5  & 30 & 0.221 $\sim$ 0.975  \\
            1 & 2  & 18 & 0.050 $\sim$ 0.975  \\
            1 & 2  & 30 & 0.122 $\sim$ 0.975  \\
            1 & 1  & 18 & 0.050 $\sim$ 0.975  \\
            1 & 1  & 30 & 0.122 $\sim$ 0.975  \\
            1 & 1  & 18 & 0.050 $\sim$ 0.975  \\
            2 & 10 & 10 & 0.050 $\sim$ 0.975  \\
            2 & 10 & 30 & 0.221 $\sim$ 0.975  \\
            2 & 5  & 14 & 0.050 $\sim$ 0.975  \\
            2 & 5  & 30 & 0.122 $\sim$ 0.975  \\
            \bottomrule
        \end{tabular}
    \end{minipage}
    \hfill
    \begin{minipage}[t]{0.32\textwidth}
        \centering
        \begin{tabular}{lrcc}
            \toprule
            Mimic & $N$ & $d_1$ (\si{cm}) & $U$ (\si{m/s}) \\
            \midrule
            2 & 2  & 18 & 0.050 $\sim$ 0.975  \\
            2 & 2  & 30 & 0.091 $\sim$ 0.975  \\
            2 & 1  & 22 & 0.050 $\sim$ 0.975  \\
            2 & 1  & 30 & 0.091 $\sim$ 0.975  \\
            3 & 10 & 14 & 0.050 $\sim$ 0.975  \\
            3 & 10 & 30 & 0.122 $\sim$ 0.975  \\
            3 & 5  & 18 & 0.050 $\sim$ 0.975  \\
            3 & 5  & 30 & 0.122 $\sim$ 0.975  \\
            3 & 2  & 22 & 0.050 $\sim$ 0.975  \\
            3 & 2  & 30 & 0.091 $\sim$ 0.975  \\
            3 & 1  & 30 & 0.050 $\sim$ 0.975  \\
            4 & 10 & 10 & 0.050 $\sim$ 0.975  \\
            4 & 10 & 14 & 0.067 $\sim$ 0.975  \\
            4 & 10 & 18 & 0.091 $\sim$ 0.975  \\
            \bottomrule
        \end{tabular}
    \end{minipage}
    \hfill
    \begin{minipage}[t]{0.32\textwidth}
        \centering
        \begin{tabular}{lrcc}
            \toprule
            Mimic & $N$ & $d_1$ (\si{cm}) & $U$ (\si{m/s}) \\
            \midrule
            4 & 10 & 22 & 0.164 $\sim$ 0.975  \\
            4 & 10 & 26 & 0.221 $\sim$ 0.975  \\
            4 & 10 & 30 & 0.297 $\sim$ 0.975  \\
            4 & 5  & 14 & 0.050 $\sim$ 0.975  \\
            4 & 5  & 18 & 0.091 $\sim$ 0.975  \\
            4 & 5  & 22 & 0.122 $\sim$ 0.975  \\
            4 & 5  & 26 & 0.164 $\sim$ 0.975  \\
            4 & 5  & 30 & 0.221 $\sim$ 0.975  \\
            4 & 2  & 18 & 0.050 $\sim$ 0.975  \\
            4 & 2  & 26 & 0.091 $\sim$ 0.975  \\
            4 & 2  & 30 & 0.122 $\sim$ 0.975  \\
            4 & 1  & 22 & 0.050 $\sim$ 0.975  \\
            4 & 1  & 30 & 0.091 $\sim$ 0.975  \\
              &    &    &                     \\
            \bottomrule
        \end{tabular}
    \end{minipage}
    \label{tab:submergence_depth}
\end{table}

For each combination of current speed and blade mimic, three repeated tests were conducted.
As shown in \cref{fig:raw_data_and_time_windows}, each repetition involved a \SI{5}{min} test, with the middle \SI{3}{min} time window used for analysis of both the bulk drag coefficient and the flutter frequency, to exclude seiching in the tank or other transients caused by changes in current velocity.
The corresponding means and standard deviations of the current speed $U$ and the moment $M_y$ are reported in \cref{tab:mean_and_sd}.
The loads on the frame alone were measured in the same manner for the eleven flow speeds at all submergence depths, without the model blades.
These baseline loads (time series) were then subtracted from the corresponding tests with the model blades. The resulting data were then used for further analysis.

\begin{table}[]
\center
\caption{Representative measurements (mean $\pm$ standard deviation) of current velocity $U$ and measured moment about the $y$-axis $M_y$ for Mimic 3 and $N=10$ in one aggregate and one base case without blade mimics at $d_1=\SI{18}{cm}$.
Statistics are extracted from the time series in \cref{fig:raw_data_and_time_windows}.}
\label{tab:mean_and_sd}
\begin{tabular}{crrrr}
\toprule
\multirow{2}{*}{Time window [\si{s}]} & \multicolumn{2}{c}{$U$ [\si{m/s}]}                          & \multicolumn{2}{c}{$M_y$ [\si{N.m}]}                            \\
                                 & \multicolumn{1}{c}{With mimics} & \multicolumn{1}{c}{Without mimics} & \multicolumn{1}{c}{With mimics} & \multicolumn{1}{c}{Without mimics} \\
\midrule
$60\sim240$    & $0.120\pm0.001$  & $0.123\pm0.001$ & $1.168\pm0.016$  & $0.035\pm0.021$ \\
$360\sim540$   & $0.165\pm0.001$ & $0.165\pm0.001$ & $1.332\pm0.022$  & $0.062\pm0.016$  \\
$660\sim840$   & $0.222\pm0.001$ & $0.224\pm0.001$ & $1.471\pm0.028$  & $0.115\pm0.100$   \\
$960\sim1140$  & $0.295\pm0.002$ & $0.303\pm0.002$ & $2.005\pm0.096$  & $0.215\pm0.041$  \\
$1260\sim1440$ & $0.401\pm0.002$ & $0.406\pm0.003$ & $2.530\pm0.169$  & $0.395\pm0.035$  \\
$1560\sim1740$ & $0.535\pm0.002$ & $0.543\pm0.003$ & $3.645\pm0.213$  & $0.703\pm0.073$  \\
$1860\sim2040$ & $0.718\pm0.004$ & $0.729\pm0.003$ & $6.089\pm0.393$   & $1.217\pm0.144$   \\
$2160\sim2340$ & $0.974\pm0.004$ & $0.979\pm0.003$ & $11.564\pm0.711$ & $1.804\pm0.170$   \\
\bottomrule
\end{tabular}
\end{table}

The values or range of the four non-dimensional parameters covered in the experiments are listed in \cref{tab:non_dimensional_parameters}.
In the definition of the four non-dimensional parameters in \cref{sec:analytical_model}, $C_D$ and $C_M$ are actually unknown and not likely to be directly extracted from the experiments.
Following \citet{Luhar2011}, we assume $C_D = 1.95$, unless stated otherwise.
This value lies within values established for flat plates placed normal to a  steady flow in the literature and is found to be in good agreement with experimental data, see \cref{sec:drag_reduction_static}.
The value of $C_M$ will be specified whenever it is used in the subsequent analysis.
When calculating $\upbeta$ and $\uplambda$ in the table, we assume $C_M=1$.
The mean values for the material properties, $\rho_s=\SI{1264.5}{kg/m^3}$ and $E=\SI{5.36}{MPa}$, are used.
For each blade mimic, tests were conducted at eleven different current speeds, with $N = 1, 2, 5,$ or $10$ blades in each aggregate.
In total, we carried out tests on 176 combinations of blade properties (by varying the blade mimic) and current speeds.

\begin{table}
\begin{center}
    \caption{Values or ranges of the non-dimensional parameters. In the table, $\upbeta$ is the mass ratio, $\mathrm{Ca}$ is the Cauchy number, $\mathrm{B}$ is the buoyancy parameter, and $\uplambda$ is the slenderness parameter.}
    \begin{tabular}{lcccc}
        \toprule
        Mimic & $\upbeta$ & $\mathrm{Ca}$        & $\mathrm{B}$      & $\uplambda$            \\
        \midrule
        1     & $[0.765,\, 0.970]$    & $[\num{5.46e2},\, \num{2.08e6}]$ & 2904.6 & 23.7            \\
        2     & $[0.867,\, 0.985]$    & $[\num{5.46e2},\, \num{2.08e6}]$ & 2904.6 & 11.8            \\
        3     & $[0.935,\, 0.993]$    & $[\num{5.46e2},\, \num{2.08e6}]$ & 2904.6 & ~5.4            \\
        4     & $[0.765,\, 0.970]$    & $[\num{6.28e1},\, \num{2.59e5}]$ & ~726.1 & 11.8            \\
        \bottomrule
    \end{tabular}
    \label{tab:non_dimensional_parameters}
\end{center}
\end{table}

\section{Numerical model}\label{sec:numerical_model}

In addition to the experiments, we employ an explicit truss-spring model~\citep{Wei2024a} to investigate the blade reconfiguration and drag loads numerically.
This model builds on the truss model originally proposed by \citet{Marichal2003} and later adopted for flexible fish-net cages by \citet{Kristiansen2012,Kristiansen2015}.
Bending moments are not included in the classic truss model, while \citet{Wei2024a} incorporated rotational springs to represent flexural stiffness.
The blade is divided into truss-like elements joined at lumped-mass nodes. Segment forces are calculated first and then distributed to the adjacent nodes. The nodal displacements and velocities are updated explicitly via Newton’s second law, and the new node positions are passed back to the segments to begin the next time step.

This numerical structural model utilizes the same load model defined in \cref{eq:drag,eq:added_mass} following \citet{Leclercq2018a}.
The main difference is that we model the structure as stretchable, with the tension calculated directly from Hooke's law.
This eliminates the need to solve the tension linear system and improves computational efficiency.
In contrast, \citet{Leclercq2018a} assumed an inextensible structure, treating the tension as an unknown.
Accordingly, our numerical model solves the two-dimensional governing equation of motion \cref{eq:dimensional_ge}, while \citet{Leclercq2018a} solved \cref{eq:non_ge} in which the tension $T$ is eliminated.
In our model, the tension $T$ in \cref{eq:dimensional_ge} is explicitly obtained using Hooke's law.
Specifically, the numerical model solves
\begin{equation}\label{eq:governing_equation_num}
\begin{split}
    &\mu\frac{\partial^2 \bm{r}}{\partial t^2} - EA \frac{\partial}{\partial s}\left[\left(1 - \left(\frac{\partial \bm{r}}{\partial s_0}\cdot\frac{\partial \bm{r}}{\partial s_0}\right)^{-1/2}\right)\frac{\partial \bm{r}}{\partial s_0}\right] + EI\left[\frac{\partial^4 \bm{r}}{\partial s^4} - \frac{\partial}{\partial s}\left(\left(\frac{\partial\bm{r}}{\partial s}\cdot\frac{\partial^3\bm{r}}{\partial s^3}\right)\frac{\partial\bm{r}}{\partial s} \right) \right]= \bm{q_d} + \bm{q_{am}} + \bm{B},
\end{split}
\end{equation}
where $s_0$ and $s$ are now the unstretched and stretched arc lengths along the structure, $\bm{q_d}$, $\bm{q_{am}}$, and $\bm{B}$ are the resistive drag, reactive force, and the effective weight in fluid as defined in \cref{sec:analytical_model}.

For computational efficiency, our numerical simulations solve \cref{eq:governing_equation_num} using the explicit truss-spring model.
It has been shown in \citet{Wei2024a} that the numerical results show excellent agreement with those of \cite{Leclercq2018}, who solved \cref{eq:non_ge} using a pseudo-Newton solver with an implicit time-stepping scheme.
This consistency can be attributed to the fact that both models employ the same hydrodynamic load formulation defined in \cref{eq:drag,eq:added_mass}.
See \citet{Wei2024a} for more details.

\begin{figure}
    \centering
    \includegraphics{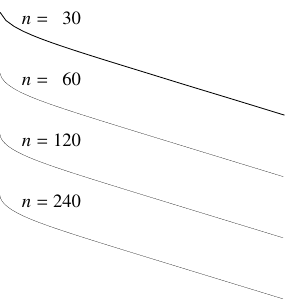}
    \hspace{0.5em}
    \includegraphics{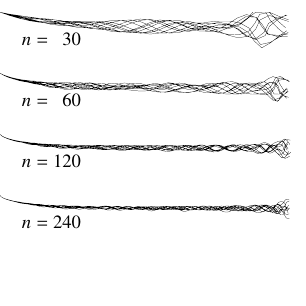}
    \caption{Convergence study for the blade trajectory as a function of the number of truss elements $n$. Snapshots of the trajectory of Mimic 2 with $N=1$. (left) In the absence of flutter, $\mathrm{Ca}=\num{3.24e4}$. (right) In the presence of flutter, $\mathrm{Ca}=\num{2.08e6}$. The number of segment $n$ is given beside the snapshots.}
    \label{fig:convergence_positions}
\end{figure}
\begin{figure}
    \centering
    \includegraphics{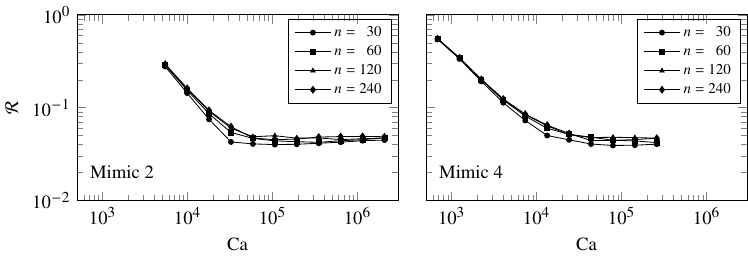}
    \caption{Convergence study of the reconfiguration number for Mimic 2 and Mimic 4 with $N=1$. The numbers of segments $n$ are given in the legend.}
    \label{fig:convergence}
\end{figure}

We carried out a convergence study for the numerical model of highly flexible blades in a steady current, both with and without flutter, see \cref{fig:convergence_positions} for an example of the snapshots of the trajectory from simulations using different numbers of truss elements.
\cref{fig:convergence} also plots the associated reconfiguration number $\mathcal{R}$ (a way to measure drag reduction, see the definition in \cref{sec:drag_reduction}) covering the experimental current velocities.
The number of segments $n$ is indicated in the legend.
In the absence of flutter, the blade’s final static shape and the drag-reduction metric converge rapidly as mesh resolution is refined.
In the presence of flutter, especially at high Cauchy numbers, the number of segments will limit the flutter wave length on the structure, as shown in \cref{fig:convergence_positions}.
The segment length constrains the shortest flutter wave length to at least four segments.
Even so, the reconfiguration number, a time-averaging and global representation of drag reduction, continues to converge, indicating that the global drag-reduction estimate is mesh-insensitive.

Due to the challenges in accurately approximating local curvature under significant deformation, particularly with short flutter waves on the structure, a fine spatial discretization is usually required.
Based on the convergence results, discretizing each blade into 120 segments is deemed sufficient for practical accuracy. This setup is used to simulate all experimental cases in this study. The numerical model follows the same strategy for equivalent thickness and bending stiffness outlined in \cref{sec:analytical_model}.
A time step of $\SI{1.0e-5}{s}$ is used to ensure numerical stability. Simulations are run for \SI{45}{s} with the last \SI{5}{s} time window used for analysis.
Each simulation takes approximately \SI{720}{s} to complete on a single Intel(R) Xeon(R) CPU E5-2660 v3@2.60 GHz.

\section{Results}\label{sec:results}
This section details our findings on the kinematic regimes of blade responses, bulk drag coefficients, and drag reduction of the flexible blades.
Given our focus on drag in the $x$-direction, we only report $F_x$ in detail.
Given the higher accuracy of $M_y$, the streamwise force $F_x$ was computed as $F_x=M_y/L$ rather than taken directly from the force measurement, where $L=\SI{0.65}{m}$ is the lever arm (the distance from the strut to the force transducer).
Accordingly, the directly measured $F_x$ was used as a consistency check.
The lifting force $F_z$ was comparable in magnitude to the drag $F_x$ and remained positive for all cases.
The remaining components $F_y$, $M_x$, and $M_z$ were near zero by symmetry.

\subsection{Static reconfiguration and flutter}
Two distinct reconfiguration regimes of blade responses were observed in the experiments.
At low values of $\mathrm{Ca}$, the blades were deflected in the direction of the flow and settled into a static position.
As $\mathrm{Ca}$ increased, the blades became more aligned with the horizontal plane.
At certain critical values of $\mathrm{Ca}$, flutter was observed to initiate from the blade tip.
As $\mathrm{Ca}$ continued to increase, flutter responses propagated towards the root, eventually covering most of the blade length.
Example photographs of the static reconfiguration and flutter are shown in \cref{fig:two_regimes}.
Several videos recorded during the experiments are also available in the supplementary material.
Snapshots from our numerical model illustrating the transition from the static reconfiguration to flutter for Mimic 2 with $C_M=1.0$ and $N=1$ are provided in \cref{fig:static_to_flutter}.

\begin{figure}
    \centerline{\includegraphics[width=0.7\textwidth]{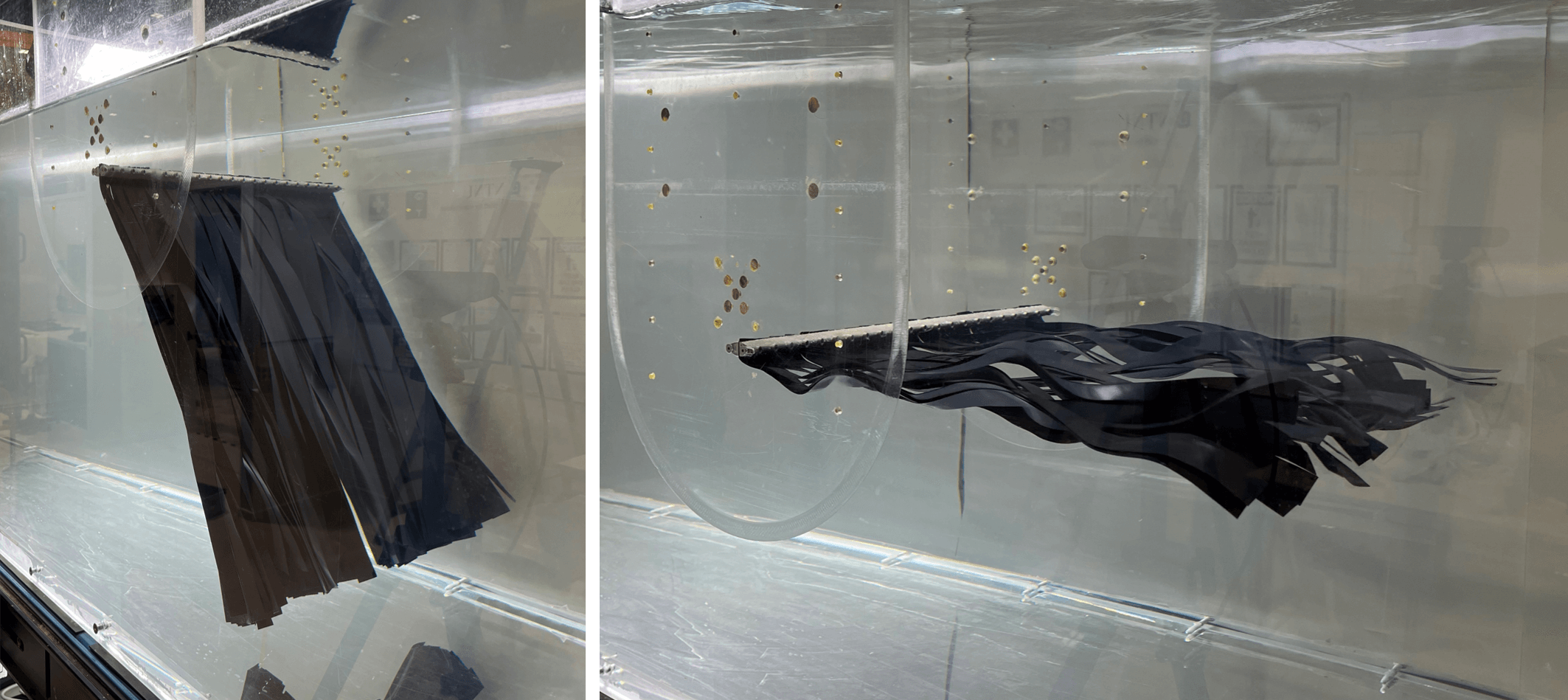}}
    \caption{The static reconfiguration at $\mathrm{Ca}=\num{9.80e2}$ (left) and flutter at $\mathrm{Ca}=\num{5.33e4}$ (right) of Mimic 1.}
    \label{fig:two_regimes}
\end{figure}

\begin{figure}
    \centerline{\includegraphics{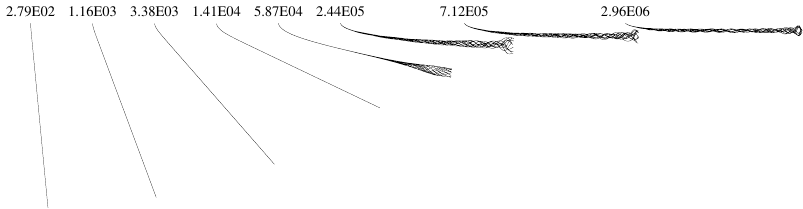}}
    \caption{Blade reconfiguration transitioning from the static to the flutter regime of Mimic 2 with $C_M=1.0$ predicted by the numerical model. The corresponding $\mathrm{Ca}$ is indicated above each image.}
    \label{fig:static_to_flutter}
\end{figure}

\begin{figure}
    \centering
    \includegraphics{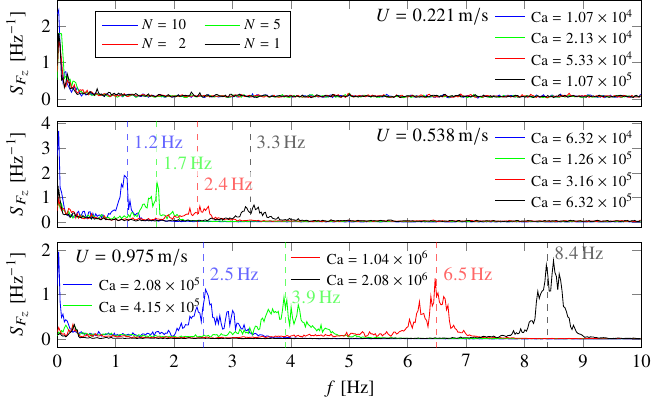}
    \caption{Normalized power spectral densities of the measured vertical force for Mimic 3 with $N=1 ,2, 5$, and $10$ at $U=\num{0.221}$, $\SI{0.538}{m/s}$, and $\SI{0.975}{m/s}$. Flutter frequencies are indicated by dashed vertical lines. The corresponding $\mathrm{Ca}$ values are annotated.}
    \label{fig:psd}
\end{figure}

Flutter is also evident in the power spectral density (PSD) of the measured vertical forces.
As an example, \cref{fig:psd} presents PSDs for Mimic 3 with $N=1,2,5$, and $10$ at $U=\num{0.221}$, $\num{0.538}$, and $\SI{0.975}{m/s}$.
Prior to PSD estimation, each segment of raw signals was subtracted from its mean value, and the resulting spectra were variance-normalized to enable shape comparison.
Apart from a peak near zero, the static regime shows no spectral peaks, e.g., $U=\SI{0.221}{m/s}$.
Once flutter develops, a pronounced peak emerges, consistent with visual observations.
With increasing ambient velocity, the flutter frequency rises, reflecting higher internal tension driven by increased drag on the blades.
At a fixed flow speed, smaller $N$ yields a higher flutter frequency.
The flutter frequencies shown in \cref{fig:psd} are also confirmed by visual observations from the experimental videos.

\begin{figure}
    \centering
    \includegraphics{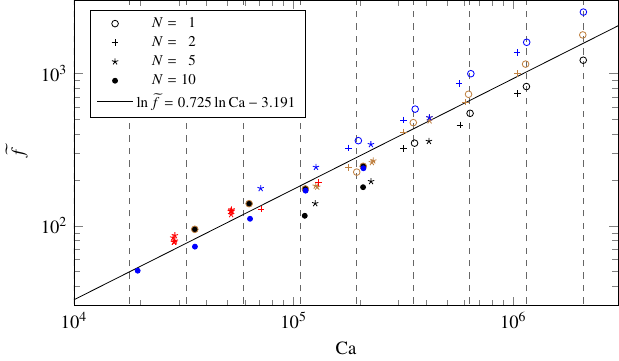}
    \caption{Normalized flutter frequencies $\widetilde{f}$ versus $\mathrm{Ca}$. Mimic 1--4 are respectively represented in brown, blue, black, and red. The number of blade per aggregate $N$ is represented by different markers: circle $\circ$, $N=1$; plus $+$, $N=2$; five-pointed star $\star$, $N=5$; dot $\bullet$, $N=10$. The vertical dashed lines represent the target values of $\mathrm{Ca} \in [\num{1.79e4}, \num{2.08e6}]$, which are evenly spaced on a logarithmic scale.}
    \label{fig:flutter_frequencies}
\end{figure}

From the PSDs computed for each test and observation, we extracted the flutter frequency $f$ for all cases in which flutter was observed.
As in \cref{sec:analytical_model}, frequencies were non-dimensionalized using $T_n$ as $\widetilde{f} = T_n f$, and the resulting values are plotted against $\mathrm{Ca}$ in \cref{fig:flutter_frequencies}.
Each marker shows the mean $\widetilde{f}$ and the corresponding $\mathrm{Ca}$ over three repetitions for a given test condition. Because the uncertainty (see \cref{app:uncertainty}) is generally small relative to the mean ($< 10\%$), error bars are omitted for clarity.
The data form clusters near the target values of $\mathrm{Ca}$.
Owing to the use of equivalent thickness and equivalent bending stiffness, the same mimic tested with different $N$ can attain the same $\mathrm{Ca}$ by adjusting the ambient velocity $U$.
Overall, the flutter frequency scales approximately as a power law with $\mathrm{Ca}$ (i.e. linear on log-log axes) and shows no systematic dependence on aspect ratio, buoyancy parameter, or mass ratio.
A linear regression $\ln \widetilde{f} = a\ln\mathrm{Ca} + b$ yields $a=0.725\pm0.047$ and $b=-3.191\pm 0.578$, where the $\pm$ values denote 95\% confidence-interval half-widths.
In more detail, Mimic~2 exhibits higher flutter frequencies than Mimic~1, and in turn than Mimic~3.
The origin of this ordering is not clear, particularly since it does not follow the aspect-ratio ranking.
Mimic~2 and Mimic~3 display similar trends, as expected, because variations in effective gravity are unlikely to provide the dominant restoring force for flutter.
The post-critical response in the flutter regime reflects strong nonlinear fluid–structure interaction, including increased local curvature and vortex shedding from the tip.
A more detailed investigation of these mechanisms is left for future work.

\subsection{Bulk drag coefficient}\label{sec:bulk_drag_coeff}
\begin{figure}
    \centerline{\includegraphics{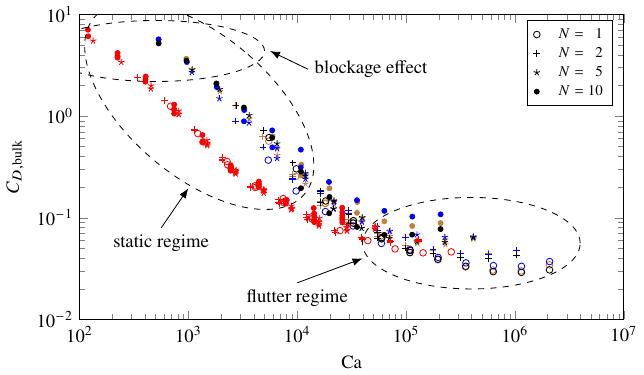}}
    \caption{The extracted $C_{D,\mathrm{bulk}}$ from the experiments versus $\mathrm{Ca}$. Mimic 1--4 are respectively represented in brown, blue, black, and red. The number of blade per aggregate $N$ is represented by different markers: circle $\circ$, $N=1$; plus $+$, $N=2$; five-pointed star $\star$, $N=5$; dot $\bullet$, $N=10$.}
    \label{fig:exp_drag_coefficients}
\end{figure}

In practical applications, the primary concern is the hydrodynamic loads acting on the blades.
The bulk drag coefficient $C_{D,\mathrm{bulk}}$~\citep{Vogel1989,Kobayashi1993} can be calculated from the mean horizontal force $\overline{F}_x=\overline{M}_y/L$ using the following expression:
\begin{equation}\label{eq:bulk_drag_coeff}
    C_{D,\mathrm{bulk}} = \frac{\overline{F}_x}{1/2 \rho b_{\mathrm{strut}} U^2 l},
\end{equation}
where $b_{\mathrm{strut}}=\SI{576}{mm}$ is the strut length, or total width of the blades, not accounting for gaps $\delta$ between the blades.
The concept of the bulk drag coefficient is practical in engineering applications. Once extracted, it can be used to estimate the drag force based on blade dimensions and the ambient current speed.
The extracted values of $C_{D,\mathrm{bulk}}$ from 303 tests are plotted in \cref{fig:exp_drag_coefficients}.
Error bars have been omitted in this figure for clarity.
The uncertainty analysis is provided in \cref{app:uncertainty}.
The discrepancies observed for some mimics at the same $\mathrm{Ca}$ and $N$ arise from using different submergence depths while holding all other conditions unchanged. This was done to prevent the blades from touching the tank bottom and to minimize the effects of the free surface.
The bulk drag coefficients agree well with those obtained from the measured $F_x$ with $R^2=0.989$.

In the static regime, $C_{D,\mathrm{bulk}}$ monotonically decreases with increasing flow speed.
Around the critical $\mathrm{Ca}$ value at which flutter occurs, a noticeable change is observed.
In the flutter regime, $C_{D,\mathrm{bulk}}$ becomes nearly constant.
This general trend aligns with the findings reported by \citet{Buck2005}.
At small $\mathrm{Ca}$ within ranges $[\num{1e2},\, \num{2e3}]$, some $C_{D,\mathrm{bulk}}$ values exceed $C_D=1.95$, which could not happen in an infinite fluid domain.
As discussed in \cref{app:blockage_effect}, a blockage effect due to the limited fluid domain bounded by the free surface and the tank floor is the cause of this.
The blockage effect is further investigated in \cref{app:blockage_effect}.
In this section, we discuss the effects of the non-dimensional parameters.

\subsubsection{Parametric analysis}
\begin{figure}
    \centerline{\includegraphics{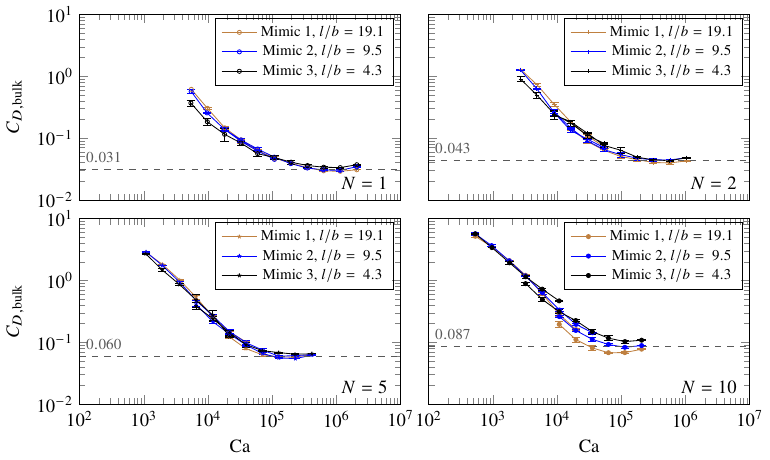}}
    \caption{Effect of the aspect ratio $l/b$ on $C_{D,\mathrm{bulk}}$ for Mimics 1--3 of the same $\mathrm{B}$ with different numbers of blades per aggregate ($N$). The dashed horizontal lines denote the average $C_{D,\mathrm{bulk}}$ in the flutter regime from the three mimics with varying aspect ratios.}
    \label{fig:effects_aspect_ratio}
\end{figure}

The effect of $l/b$ instead of $\uplambda$ is discussed first because $l/b$ is known, whereas $\uplambda$ depends on unknown $C_D$ and $C_M$.
Mimic 1, 2, and 3 share the same buoyancy parameter and differ in aspect ratio, see \cref{tab:mimic_dimensions} and \cref{tab:non_dimensional_parameters}.
As shown in \cref{fig:effects_aspect_ratio}, varying the aspect ratio generally has a small effect on $C_{D,\mathrm{bulk}}$ in both regimes, nor does it significantly affect the critical $\mathrm{Ca}$.
This finding contrasts with the conclusions drawn for individual blades by \citet{Leclercq2018a}.
In their numerical study, both $C_D$ and $C_M$ were fixed, allowing $l/b$ to effectively represent $\uplambda$.
They found that increasing the length-to-width ratio $l/b$ stabilized the system.
The more slender the structure, i.e., larger $l/b$, the higher the critical $\mathrm{Ca}$ required for flutter onset.
The findings of \citet{Leclercq2018a} are based on numerical simulations and, to our knowledge, have not yet been confirmed experimentally.
As discussed in \cref{sec:limitations}, models developed for individual blades may not accurately represent the behavior of side-by-side blades.
Our numerical results with fixed $C_D$ and $C_M$ also confirm this, as will be discussed later in association with \cref{fig:exp_num_drag_reduction} in \cref{sec:limitations}.

\begin{figure}
    \centerline{\includegraphics{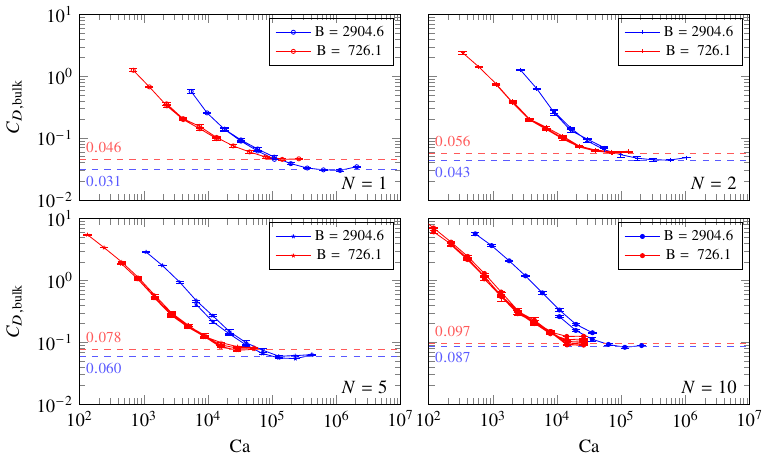}}
    \caption{The effect of the buoyancy parameter $\mathrm{B}$ on $C_{D,\mathrm{bulk}}$ between Mimic 2 and Mimic 4 of the same aspect ratio $l/b$ with different numbers of blades per aggregate ($N$). The dashed horizontal lines denote $C_{D,\mathrm{bulk}}$ in the flutter regime.}
    \label{fig:effects_buoyancy}
\end{figure}

Mimic 2 and Mimic 4 share the same $l/b$ but differ in the buoyancy parameter $\mathrm{B}$.
As demonstrated in \cref{fig:effects_buoyancy}, in the static regime, a larger buoyancy parameter $\mathrm{B}$ results in a higher $C_{D,\mathrm{bulk}}$, as the effective weight $\bm{B}$ pulls the blades downward, increasing the projected area against the flow.
As a restoring force, the effective weight $\bm{B}$ stabilizes the system by delaying the onset of flutter.
By delay, it means the flutter occurs at a higher $\mathrm{Ca}$.

When the number of blades in an aggregate is increased, the buoyancy parameter $\mathrm{B}$ and the length-to-width ratio $l/b$ remain unchanged.
However, since $C_M$ appears in both $\upbeta$ and $\uplambda$, and $C_D$ is included in $\uplambda$ as well, it becomes challenging to isolate the effects of $N$ on $C_D$ and $C_M$, and subsequently to assess the impact of $\upbeta$ on $C_{D,\mathrm{bulk}}$.
The interplay between these parameters complicates directly evaluating how blade number affects the overall drag coefficient.
Nevertheless, increasing $N$ should reduce $\upbeta$. In other words, as the body mass increases with more blades in an aggregate, the proportion of the fluid inertia in the system decreases.

\begin{figure}
    \centerline{\includegraphics{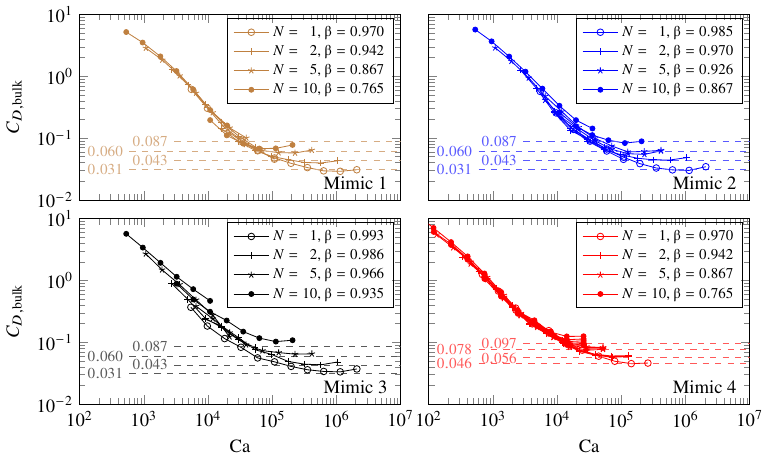}}
    \caption{The effect of the mass ratio $\upbeta$ on $C_{D,\mathrm{bulk}}$ for Mimics 1--4. $\upbeta$ is computed with $C_M=1.0$. The dashed horizontal lines denote $C_{D,\mathrm{bulk}}$ in the flutter regime. For $N=10$, we use the average $C_{D,\mathrm{bulk}}=0.087$ from Mimic 1--3, see \cref{fig:effects_aspect_ratio}.
    For clarity, the error bars are not plotted.}
    \label{fig:effects_mass_ratio}
\end{figure}

Observations (\cref{fig:effects_mass_ratio}) indicate that the effect of $\upbeta$ on $C_{D,\mathrm{bulk}}$  is minor for all four mimics in the static regime.
However, $\upbeta$ does have a stabilizing effect and thus a larger $\upbeta$ leads to a higher critical $\mathrm{Ca}$ where flutter starts to develop.
A main effect of $\mathrm{\upbeta}$ is to reduce $C_{D,\mathrm{bulk}}$ in the flutter regime as it tends to unity.
Our experiments show a factor as large as $2-3$.

In the analysis, we have assumed constant values of $C_D=1.95$ and $C_M$, which are included in the definition of $\mathrm{Ca}$.
An alternative definition of $\mathrm{Ca}$ without $C_D$, expressed as
\begin{equation}
    \mathrm{Ca}' = \frac{1}{2}\frac{\rho b U^2 l^3}{EI},
\end{equation}
is also widely used in the literature~\citep{Gosselin2010,Jacobsen2019a,Vettori2024}.
Using this alternative definition, the previous qualitative conclusions still hold. However, the curves plotted on logarithmic scales would be shifted a certain distance to the left.

\subsubsection{Analytical prediction in the static regime}

\begin{figure}
    \centerline{\includegraphics{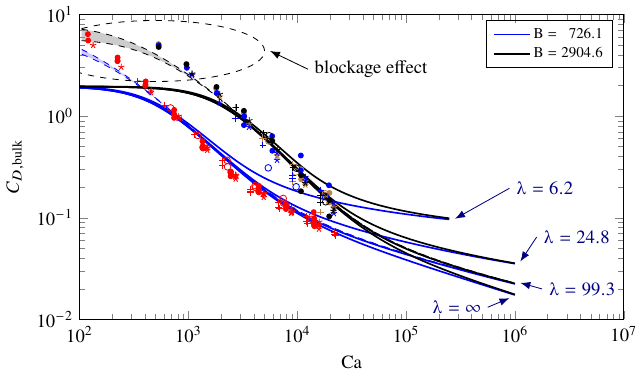}}
    \caption{The analytical predictions of $C_{D,\mathrm{bulk}}$ with different $\uplambda$ (lines) and the experimental data (markers) in the static regime. The dashed lines are analytical predictions of $C_{D,\mathrm{bulk}}$ accounting for the blockage effect. The shaded areas between the dashed lines are bounded by $\delta/b=0$ and $0.05$. Mimic 1--4 are respectively represented in brown, blue, black, and red. The number of blade per aggregate $N$ is represented by different markers: circle $\circ$, $N=1$; plus $+$, $N=2$; five-pointed star $\star$, $N=5$; dot $\bullet$, $N=10$. For clarity, some data in the flutter regime have been removed in this plot.}
    \label{fig:exp_ana_drag_reduction}
\end{figure}

Having obtained the reconfiguration number $\mathcal{R}$ from the analytical model in \cref{sec:drag_reduction}, the corresponding bulk drag coefficient follows $C_{D,\mathrm{bulk}}=C_D \mathcal{R}$ for a given $C_D$.
The analytical predictions of $C_{D,\mathrm{bulk}}$ with $C_D=1.95$ are plotted together with the experimental data in the static regime for $\mathrm{Ca} \le 2\times 10^4$ in \cref{fig:exp_ana_drag_reduction}.

Using the analytical model, we can examine the effect of $\uplambda$.
With the same $\mathrm{B}$, all curves for different $\uplambda$ coincide at small $\mathrm{Ca}$ until the curve for smaller $\uplambda$ changes its rate of decrease and starts to deviate.
As $\uplambda$ increases, the curve approaches the solution without the reactive force term in \cref{eq:non_static_ge} by letting $\uplambda\to\infty$.
From a physical standpoint, the reactive force disappears in the static regime because there is no relative acceleration between the structure and the surrounding fluid, and the added mass in \cref{eq:added_mass} does not apply.
Accordingly, the limit $\uplambda\to \infty$ produces the closest agreement with the experimental data in the static regime $\mathrm{Ca} \le 2\times 10^4$.

The mismatch at very low $\mathrm{Ca}$ ($\mathrm{Ca}<700$ for $\mathrm{B}=726.1$ and $\mathrm{Ca}<2000$ for $\mathrm{B}=2904.6$) in \cref{fig:exp_ana_drag_reduction} stems from tank-blockage effects. For small $\mathrm{Ca}$, the blades exhibit less deformation, leading to a blockage of the current due to the limited depth of the tank and an increase in ambient current velocity. Once this blockage is accounted for (see \cref{app:blockage_effect}), the analytical predictions based on the solutions of \cref{eq:non_static_ge} agree with the experimental results.
The blockage effect also explains why $F_x$ with blade mimics does not converge to zero as $U$ goes to zero in \cref{fig:raw_data_and_time_windows}.
The same behavior appears in other cases.
As $U$ decreases, the blockage effect becomes more significant, $C_{D,\mathrm{bulk}}$ increases above $2$, even up to 8 (see \cref{fig:exp_drag_coefficients,fig:exp_ana_drag_reduction}).
Consequently, $F_x$ is not quadratic in $U$ anymore and a simple back-extrapolation of $F_x$ versus $U$ would not pass through the origin.

\begin{figure}
    \centering
    \includegraphics{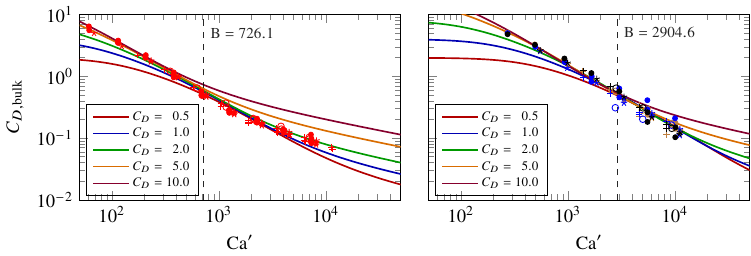}
    \caption{The analytical predictions of $C_{D,\mathrm{bulk}}$ with different values of $C_D$ and the experimental data (markers) in the static regime. The vertical dashed line denotes the corresponding buoyancy parameter $\mathrm{B}$ in each plot. We let $\uplambda\to\infty$ and $\delta/b=0$.}
    \label{fig:exp_ana_bulk_drag_static_varying_CD}
\end{figure}

So far, we have used $C_D=1.95$ following \citet{Luhar2011}, and it provides good agreement with the experimental data, as shown in \cref{fig:exp_ana_drag_reduction}.
To identify the optimal value of $C_D$, we compare the analytical prediction of $C_{D,\mathrm{bulk}}$ for different values of $C_D$ with the experimental data in \cref{fig:exp_ana_bulk_drag_static_varying_CD}.
The blockage effect is included, and we use $\uplambda\to\infty$ and $\delta/b=0$.
In the range $\mathrm{Ca}/B > 1$, where blockage is minimal, $C_D=2$ provides the best fit.
In the range $\mathrm{Ca}/B < 1$, where blockage becomes significant, $C_D=2$ still shows good agreement.
In the flutter regime, asynchronous motion increases the spacing between aggregates, so $C_D=2.0$ for rigid plates remains applicable.
Overall, the use $C_D=1.95$ following \citet{Luhar2011} remains a reasonable choice.

In conclusion, the analytical model is capable of predicting drag reduction in the static regime. At small $\mathrm{Ca}$, if the blockage effect is significant, the estimate should be adjusted based on the solidity ratio as a consequence of the limited tank size.
The numerical model also estimates drag reduction across both static and flutter regimes: the predicted values of $C_{D,\mathrm{bulk}}$ from Section \ref{sec:bulk_drag_coeff} are rescaled by the cross-sectional $C_D$ coefficients employed in the simulations.

\subsubsection{Numerical prediction in both regimes}

\begin{figure}
    \centerline{\includegraphics{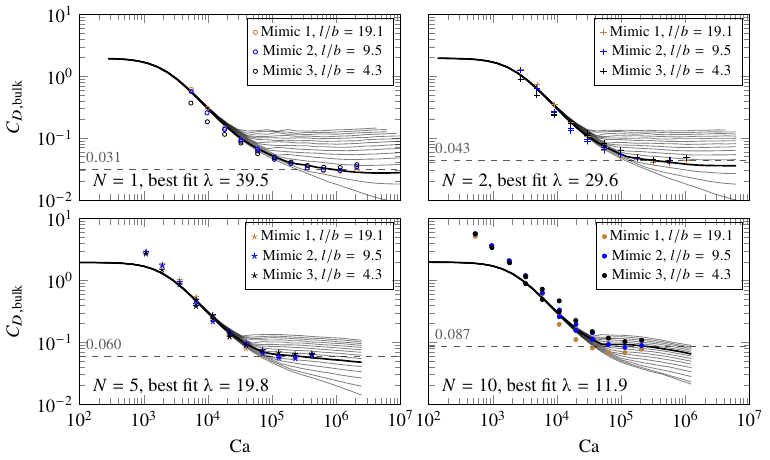}}
    \caption{The numerical predictions of the bulk drag coefficient $C_{D,\mathrm{bulk}}$ with different $\uplambda$ values (lines) and the experimental data (markers). The thick curve denotes the best-fit $\uplambda$.
    }
    \label{fig:exp_num_drag_reduction}
\end{figure}

We have seen in our experiments that $l/b$ has minimal influence on $C_{D,\mathrm{bulk}}$ in both the static and flutter regimes, as discussed previously.
We now proceed to investigate the effect of $\uplambda$ using the numerical model described in \cref{sec:numerical_model}.
We fix $C_D=1.95$ and adjust $C_M$ from $0.10$ up to $3.00$ to change $\uplambda$ for Mimic 1--3.
Selected numerical predictions of $C_{D,\mathrm{bulk}}$ for Mimic 1 ($C_M=0.2,0.4,\dots,3.0$ and $\uplambda=118.6,59.3,\dots,7.9$) are compared to the experimental data in \cref{fig:exp_num_drag_reduction}.
For Mimic 1--3 (same $\mathrm{B}$, different $l/b$), numerical cases with equal $\uplambda$ by varying $C_M$ yield identical results; numerical predictions for Mimics 2 and 3 are not shown here for brevity.
In the static regime, the numerical predictions closely match the experimental data for all $\uplambda$.
However, $\uplambda$ plays a significant role in determining both the critical $\mathrm{Ca}$ and $C_{D,\mathrm{bulk}}$ in the flutter regime.
A larger $\uplambda$ delays the onset of flutter and results in a smaller $C_{D,\mathrm{bulk}}$ within the flutter regime.
In the limit of $\uplambda \to \infty$, no flutter occurs.
To replicate the flutter observed in the experiments, $\uplambda$ must be retained, and the terms involving $\uplambda$, i.e., the last two terms in \cref{eq:added_mass} must also be included in the hydrodynamic load model.
This point is also confirmed from a different perspective in \cref{sec:reactive_force_model}.

With specific values of $C_M$ and $\uplambda$, our numerical model is able to predict both the critical $\mathrm{Ca}$ and $C_{D,\mathrm{bulk}}$ in the flutter regime with good fidelity.
For Mimics 1--3, which have different $l/b$, similar values of $\uplambda$ are required to align numerical predictions of $C_{D,\mathrm{bulk}}$ with experimental data, as summarized in \cref{tab:best_fit_values}.
By the definition of $\uplambda$, this implies that $C_D/C_M$ must be inversely proportional to $l/b$ to maintain the same $\uplambda$.
Based on the assumption that $C_D$ is fixed, it follows that $C_M$ is proportional to $l/b$.
This also aligns with the dimensional analysis in \cref{sec:governing_equations}.
When $\uplambda$ is matched by varying $C_M$, the numerical results for Mimic 1--3 are identical because the remaining non-dimensional parameters remain unchanged across mimics.
$\mathrm{Ca}$ and $\mathrm{B}$ do not change since they are independent of $C_M$.
From \cref{eq:mass_ratio_N_layers}, $\upbeta$ does not change if $C_M \propto l/b$ with $l$ fixed.
The conclusion here is based on the limited range of $l/b$ in our experiments.
Further studies are needed to examine its generality.

Since the numerical solver simulates the fully nonlinear governing equation of blade motion and uses a complete load model, it can adequately represent the physics, provided that accurate $C_D$ and $C_M$ are applied.
We conclude that, for side-by-side blades, the cross-sectional hydrodynamic coefficients $C_D$ and $C_M$ are influenced by the blade's aspect ratio $l/b$ such that $\uplambda = 2/\pi(C_D/C_M) (l/b)$ remains nearly constant.

\begin{table}
    \center
    \caption{Best-fit values of $\uplambda$ and $C_M$ for Mimic 1--3. A constant $C_D=1.95$ was used for all three mimics. The buoyancy parameter $\mathrm{B}$, length-to-width ratio $l/b$, and mass ratio $\upbeta$ are also provided.}
    \begin{tabular}{lcccccccccccccc}
    \toprule
    \multirow{2}{*}{Mimic} &
      \multirow{2}{*}{$\mathrm{B}$} &
      \multirow{2}{*}{$l/b$} &
      \multicolumn{3}{c}{$N=1$} &
      \multicolumn{3}{c}{$N=2$} &
      \multicolumn{3}{c}{$N=5$} &
      \multicolumn{3}{c}{$N=10$} \\
     &
       &
       &
      $C_M$ &
      $\uplambda$ &
      $\upbeta$ &
      $C_M$ &
      $\uplambda$ &
      $\upbeta$ &
      $C_M$ &
      $\uplambda$ &
      $\upbeta$ &
      $C_M$ &
      $\uplambda$ &
      $\upbeta$ \\
      \midrule
        1 & 2904.6 & 19.1  & 0.60 & 39.5 & 0.951 & 0.80 & 29.6 & 0.929 & 1.20 & 19.8 & 0.887 & 2.00 & 11.9 & 0.867 \\
        2 & 2904.6 & ~~9.5 & 0.30 & 39.3 & 0.951 & 0.40 & 29.5 & 0.929 & 0.60 & 19.7 & 0.887 & 1.00 & 11.8 & 0.867 \\
        3 & 2904.6 & ~~4.3 & 0.14 & 38.1 & 0.952 & 0.18 & 29.7 & 0.928 & 0.27 & 19.8 & 0.885 & 0.45 & 11.9 & 0.866 \\
        \bottomrule
    \end{tabular}
    \label{tab:best_fit_values}
\end{table}

From these results, we argue that our numerical model has sufficient physics incorporated to capture the onset of flutter, and even to predict the (nearly $\mathrm{Ca}$-independent) $C_{D,\mathrm{bulk}}$ within the flutter regime, as evidenced by the curves with varying $\uplambda$ as long as appropriate $C_D$ and $C_M$ are provided.

The effects of the non-dimensional parameters on $C_{D,\mathrm{bulk}}$ in both regimes and system stability from the experimental and numerical results are summarized in \cref{tab:effects_non_dimensionals}.
These effects are consistent with the general trend shown in \cref{fig:exp_drag_coefficients}.
In the static regime, $C_{D,\mathrm{bulk}}$ decreases monotonically until flutter occurs, after which it transitions to an almost constant value.
For example, $\upbeta$ has no effect on $C_{D,\mathrm{bulk}}$ in the static regime but plays a stabilizing role. As a result, a larger $\upbeta$ will delay the onset of flutter, which in turn leads to a lower $C_{D,\mathrm{bulk}}$ in the flutter regime.
The effects of $\mathrm{B}$ are more complicated. $\mathrm{B}$ and $C_{D,\mathrm{bulk}}$ are positively correlated in the static regime and meanwhile positive $\mathrm{B}$ stabilizes the system and delays the flutter. As a result, it is difficult to predict how $C_{D,\mathrm{bulk}}$ in the flutter regime will change with variations in $\mathrm{B}$, which may vary at different values of $\upbeta$ or $\uplambda$.

The overall trend of $C_{D,\mathrm{bulk}}$ for side-by-side blades resembles that of a single blade in a uniform current~\citep{Leclercq2018a} with some differences.
One is that in the flutter regime, $C_{D,\mathrm{bulk}}$ for side-by-side blades levels off at a constant value, while for a single blade, it continues to decrease, although at a slower rate than in the static regime, as noted by \citet{Leclercq2018a}.

\begin{table}
\begin{center}
    \caption{Effects of the non-dimensional parameters on $C_{D,\mathrm{bulk}}$ and system stability.}
    \begin{tabular}{lccc}
        \toprule
        Term  \quad   &\quad $C_{D,\mathrm{bulk}}$ in static regime \quad & \quad System stability \quad & \quad $C_{D,\mathrm{bulk}}$ in flutter regime \\
        \midrule
        $\upbeta$   & none                & stabilizing    &  negatively correlated  \\
        $\mathrm{B}$        & positively correlated & stabilizing    & depends on $\upbeta$ and $\uplambda$  \\
        $l/b$      & none                & none             & none         \\
        $\uplambda$ & none &  stabilizing & negatively correlated \\
        \bottomrule
    \end{tabular}
    \label{tab:effects_non_dimensionals}
\end{center}
\end{table}

\section{Discussion}\label{sec:discussion}
In this section, we first revisit the reactive force model to identify which terms are responsible for triggering flutter responses.
We then discuss the applicability of the hydrodynamic force model to our experimental setup.
Finally, we discuss the limitations in our study and offer suggestions for future research.

\subsection{Revisiting the reactive force model}\label{sec:reactive_force_model}

\begin{figure}
    \centerline{\includegraphics{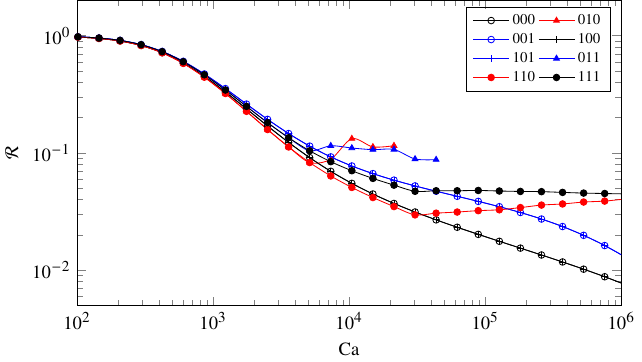}}
    \caption{The numerical prediction of $\mathcal{R}$ for Mimic 4 with different combinations of the three terms in the reactive force model. The three codes represent $\bm{q}_{\bm{am},2,1}$, $\bm{q}_{\bm{am},2,2}$, and $\bm{q}_{\bm{am},3}$, respectively. 1 means the corresponding term is included, and 0 means that term is excluded.}
    \label{fig:mimic_4_term_by_term}
\end{figure}

By testing eight different combinations of including or excluding the terms $\bm{q}_{\bm{am},2,1}$, $\bm{q}_{\bm{am},2,2}$, and $\bm{q}_{\bm{am},3}$ in the hydrodynamic load model, the corresponding $\mathcal{R}^{\mathrm{num}}$ are plotted in \cref{fig:mimic_4_term_by_term}.
A three-digit code denotes the combinations of $\bm{q}_{\bm{am},2,1}$, $\bm{q}_{\bm{am},2,2}$, and $\bm{q}_{\bm{am},3}$, where $1$ means the term is activated, and $0$ means it is deactivated.
For example, Code 110 means $\bm{q}_{\bm{am},2,1}$ and  $\bm{q}_{\bm{am},2,2}$ are included but $\bm{q}_{\bm{am},3}$ is excluded.
Upon analyzing each term, we find the following effects:
\begin{itemize}
    \item
The term $\bm{q}_{\bm{am},2,2}$ has a destabilizing effect.
When this term is included (e.g., Codes 111 and 110), flutter occurs.
Conversely, when this term is excluded (e.g., Codes 101 and 100), no flutter is observed. This indicates that $\bm{q}_{\bm{am},2,2}$ contributes to system instability.
\item The term $\bm{q}_{\bm{am},2,1}$ has a stabilizing effect.
Including $\bm{q}_{\bm{am},2,1}$ alongside $\bm{q}_{\bm{am},2,2}$ (e.g., Codes 111 and 110) allows flutter to occur without causing the simulation to crash.
As $\mathrm{Ca}$ increased, the numerical solver failed to capture the shorter flutter wavelength and higher frequency.
Refining the spatial discretization or decreasing the time step size postpones the crash to higher $\mathrm{Ca}$.
However, excluding $\bm{q}_{\bm{am},2,1}$ while including $\bm{q}_{\bm{am},2,2}$ (e.g., Codes 011 and 010) leads to simulation crashes.
This situation does not occur physically, since $\bm{q}_{\bm{am},2,1}$ and $\bm{q}_{\bm{am},2,2}$ should always appear together, but it is useful here to identify which term triggers the instability.
The term $-\widetilde{U}\partial\theta/\partial\tilde{t}$ exists in both $\bm{q}_{\bm{am},1}$ and $\bm{q}_{\bm{am},2}$.
Later we can see that excluding $\bm{q}_{\bm{am},2,1}$ effectively corresponds to reducing $\upbeta$ to one quarter of its original value while maintaining the original load model.
This adjustment significantly lowers the critical $\mathrm{Ca}$, as illustrated in Figure 3 of \citet{Leclercq2018a}.
In scenarios where $\bm{q}_{\bm{am},2,2}$ is excluded and no flutter occurs, the transient term $\bm{q}_{\bm{am},2,1}$ diminishes and has no effect on $\mathcal{R}$ due to the static reconfiguration at all $\mathrm{Ca}$ (e.g., comparison between Codes 000 and 100, Codes 001 and 101).
\item The term $\bm{q}_{\bm{am},3}$, included or not, will not prevent flutter or alter its onset, but modifies $\mathcal{R}$ (e.g., comparison between Codes 111 and 110, Codes 000 and 001).
\end{itemize}

Physically speaking, $\bm{q}_{\bm{am},2,1}$ and $\bm{q}_{\bm{am},2,2}$ come from the same term $\bm{q}_{\bm{am},2}$ and should not be separated from each other. If we check $\bm{q}_{\bm{am},2}$ as a whole (e.g., Codes 001 vs. 111, Codes 000 vs. 110), this term in total contributes to the system instability.
The effects of different combinations of the reactive force components are summarized in \cref{tab:reactive_force_terms}.
In conclusion, all terms in the reactive force model matter and should be included in order to predict the onset of flutter as well as the resulting drag loads.
The term $\bm{q}_{\bm{am},1}$ alongside the drag is insufficient to trigger flutter or predict drag reduction in the application of highly compliant structures.
At low $\mathrm{Ca}$, when the blades are perpendicular to the inflow, there is mainly cross-flow separation and wake formation along the length of the blades, which is sufficiently represented by the drag term.
At high $\mathrm{Ca}$, the structure aligns with the flow, and the hydrodynamic load model represented by \cref{eq:drag,eq:added_mass} is more appropriate.

\begin{table}
\begin{center}
    \caption{Different combinations of the reactive force components and their effects on system instability.}
    \begin{tabular}{lcccc}
        \toprule
        Code & 000 & 001 & 110 & 111 \\
        \midrule
        $\bm{q}_{\bm{am},2,1}$     & $-$          & $-$         & $+$               & $+$    \\
        $\bm{q}_{\bm{am},2,2}$     & $-$          & $-$         & $+$               & $+$    \\
        $\bm{q}_{\bm{am},3}$     & $-$          & $+$         & $-$               & $+$    \\
        System instability     & No flutter          & No flutter         & Flutter               & Flutter   \\
        \bottomrule
    \end{tabular}
    \label{tab:reactive_force_terms}
\end{center}
\end{table}

\begin{figure}
    \centerline{\includegraphics{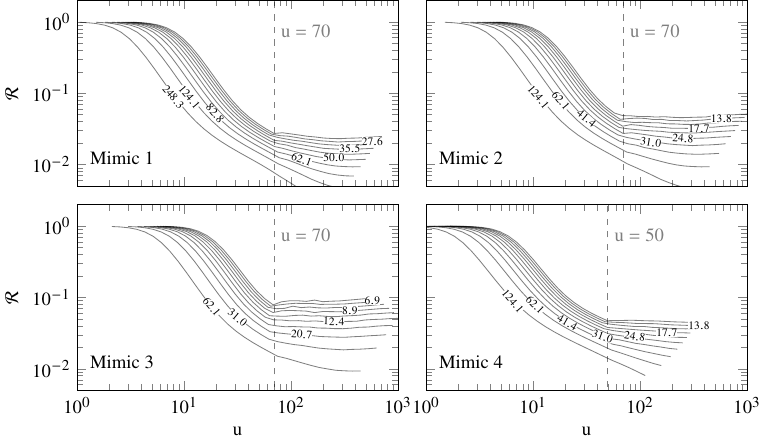}}
    \caption{The numerical predictions of drag reduction $\mathcal{R}$ with different $\uplambda$ as a function of reduced velocity $\mathrm{u}$. The vertical dashed line denotes the critical $\mathrm{u}$, where flutter occurs in the numerical simulation. Selected $\uplambda$ values are annotated on or beside the lines.}
    \label{fig:reduced_u_exp_num_drag_reduction}
\end{figure}

If we plot the numerical predictions of $\mathcal{R}$ with respect to $\mathrm{u}$ instead of $\mathrm{Ca}$ as shown in \cref{fig:reduced_u_exp_num_drag_reduction}, the regime transition points will be aligned at approximately the same $\mathrm{u}$ ($\mathrm{u}\approx 70$ for Mimics 1--3 with $\mathrm{B}=2904.6$ and $\mathrm{u}\approx 50$ for Mimic 4 with $\mathrm{B}=725.1$), or the critical $\mathrm{u}$ depends little on $\uplambda$, which again is consistent with \citet{Leclercq2018a}. Meanwhile, $\mathcal{R}$ in the static regime does not collapse onto a single curve as we have seen in \cref{fig:exp_num_drag_reduction}.
The dependence of drag reduction and flutter instability on $\mathrm{Ca}$ and $\mathrm{u}$ is summarized in \cref{tab:Ca_and_u}.
Since this paper focuses on drag reduction, most of the results have been presented as functions of $\mathrm{Ca}$.

\begin{table}
\begin{center}
    \caption{The dependence of drag reduction and flutter instability on $\mathrm{Ca}$ and $\mathrm{u}$.}
    \begin{tabular}{lcc}
        \toprule
        Regime & Cauchy number $\mathrm{Ca}$ & Reduced velocity $\mathrm{u}$ \\
        \midrule
        Static regime & Same curve & Separate curves \\
        Transition & Different critical $\mathrm{Ca}$ & Same critical $\mathrm{u}$ \\
        Flutter regime & Higher critical $\mathrm{Ca}$, lower $\mathcal{R}$ & No common rule \\
        \bottomrule
    \end{tabular}
    \label{tab:Ca_and_u}
\end{center}
\end{table}

\begin{figure}
    \centerline{\includegraphics{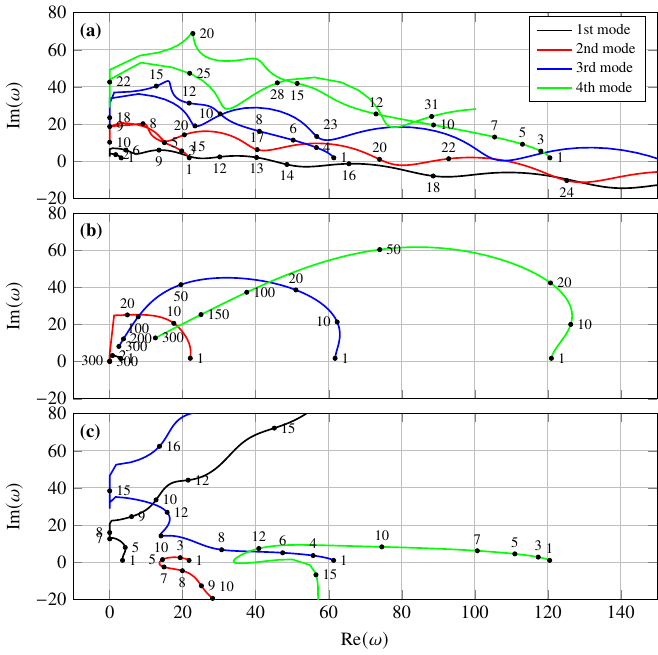}}
    \caption{The dimensionless complex frequency of the four lowest modes of the system ($\upbeta=0.8$) as a function of the reduced velocity annotated along the curves. (a) All terms are included; (b) The third term in \cref{eq:linear_beam_flutter} is excluded; (c) The second term in \cref{eq:linear_beam_flutter} is reduced to its half.}
    \label{fig:stability}
\end{figure}

We set $\upbeta=0.8$, which falls within the range of our experiments (see \cref{tab:non_dimensional_parameters}).
Following the method of \citet{Paidoussis1998}, we apply the Galerkin method using the first twelve modes in the Galerkin expansion to solve \cref{eq:linear_beam_flutter}.
The dimensionless complex frequencies of the four lowest modes are plotted in \cref{fig:stability}. Although not shown, the remaining modes were also examined to determine system stability.
A negative imaginary part of the complex frequency, i.e., $\mathrm{Im}(\omega)$ below zero in \cref{fig:stability}, means instability.
When all terms are included, instability initiates in the first mode at $\mathrm{u}\approx 13.5$.
The critical reduced velocity here is lower than $\mathrm{u}\approx 50\text{ or }70$ in \cref{fig:reduced_u_exp_num_drag_reduction}. One possible explanation is the absence of resistive drag in \cref{eq:linear_beam_flutter}.
Physically, additional damping due to drag would delay the onset of flutter to larger $\mathrm{u}$.
As noted by \citet{Leclercq2018a}, the persistent drag term in our configuration introduces some additional damping that stabilizes the system compared with the axial configuration.
As mentioned earlier in \cref{sec:analytical_model}, the drag is needed to bend the blade from the initial position perpendicular to the flow into a more aligned configuration, which is prone to flutter.
If the third term in \cref{eq:linear_beam_flutter}, the counterpart of  $\bm{q}_{\bm{am},2,2}$, is disabled, all modes remain stable regardless of the value of $\mathrm{u}$.
In this scenario, the first mode rapidly approaches the origin and diminishes.
If the second term in \cref{eq:linear_beam_flutter} is reduced by half, effectively deactivating the counterpart of $\bm{q}_{\bm{am},2,1}$, the system loses its stability in the second mode first at $\mathrm{u}\approx 6$ instead of in the first mode.
In fact, for \cref{eq:linear_beam_flutter}, halving the second term is equivalent to reducing $\upbeta$ to one quarter of its original value meanwhile solving the original form of \cref{eq:linear_beam_flutter}.
This is illustrated in \cref{fig:stability}c, which corresponds to Figure 3.30 of \citet{Paidoussis1998} where $\upbeta=0.2$ was assumed.

Overall, the critical $\mathrm{u}$ increases with $\upbeta$ with some local jumps, as shown in Figure 3.33 of \citet{Paidoussis1998} for the linear stability of \cref{eq:linear_beam_flutter} and in Figure 2 of \citet{Leclercq2018a} for scenarios involving drag, as studied in this paper.
In conclusion, our findings are consistent with these previous studies regarding the effects of the different terms in the reactive force model.

\subsection{Applicability of the hydrodynamic load model}\label{sec:applicability_hydro_load}
The hydrodynamic model used in \cref{sec:analytical_model} and \cref{sec:numerical_model} was originally developed in \citet{Lighthill1971,Candelier2011} for an individual blade situated in an infinite fluid domain.
In the following, we show that it can also be applied to our side-by-side blades in a limited fluid.

The resistive component \cref{eq:drag}, associated with flow separation in the plane of the cross sections, is essential to prevent unrealistically large flapping amplitudes in the post-critical regime following the onset of flutter~\citep{Singh2012}. Additionally, the blades in this study are initially oriented perpendicular to the flow. Resistive drag plays a crucial role in statically bending the blade within the static regime or in configuring the blade into a state prone to flutter.
Following \citet{Eloy2011,Singh2012,Leclercq2018a} and \citet{Leclercq2018}, we here refer to \cref{eq:drag} as the resistive force and \cref{eq:added_mass} as the reactive force.
Under the assumptions of the reactive force model~\citep{Lighthill1971, Candelier2011}, there is no flow separation except at the trailing edge.
The approach is similar to Morison’s decomposition, expressing the total hydrodynamic force as the sum of reactive (inertial) and resistive (drag) terms.

In the static regime, the dominant force is the resistive drag.
For very low $\mathrm{Ca} < O(\mathrm{B})$, the finite water depth plays an important role due to the blockage effect, as discussed in \cref{app:blockage_effect}.
When this effect is accounted for, the analytical model provides a good prediction of the drag in this range, as shown in \cref{fig:exp_ana_drag_reduction}.
For higher $\mathrm{Ca} > O(\mathrm{B})$, but still within the static regime, the influence of finite water depth becomes secondary.
In this range, the analytical model assuming an unbounded fluid domain accurately predicts the drag, as also shown in \cref{fig:exp_ana_drag_reduction}.

\begin{figure}
    \centerline{\includegraphics{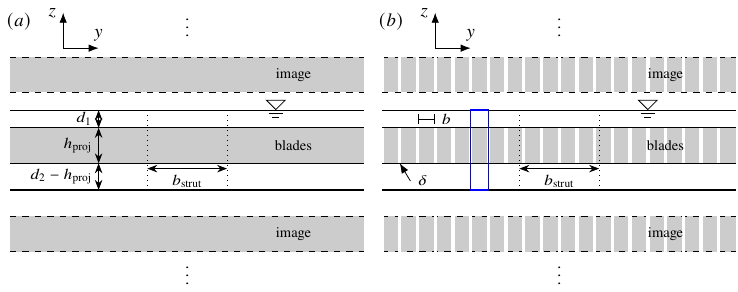}}
    \caption{(a) The blades extended in the $y$-direction and its images about the free surface and the tank bottom.  (b) Gaps between adjacent aggregates of blades. The blue box denotes the control surface surrounding an aggregate of blades.}
    \label{fig:mirroring_and_solidity_ratio_control_surface}
\end{figure}

In the flutter regime, most parts of the blades become aligned with the ambient flow, and the reactive force becomes a dominant contributor due to dynamic motion of the blades.
In our experiments, the blades were arranged side by side in a tank with limited cross-span and depth.
The free-surface disturbance due to the presence of the model is small and can be neglected.
We first mirror the setup about the free surface and the tank bottom to remove the horizontal boundaries, and replicate it periodically in the $y$-direction to remove the lateral boundaries (acrylic plates), as illustrated in \cref{fig:mirroring_and_solidity_ratio_control_surface}a.
The blade aggregates in the experiments were not perfectly aligned with each other.
As shown in \cref{fig:mirroring_and_solidity_ratio_control_surface}b, this misalignment resulted in gaps between the aggregates that allowed some flow to pass around the side edges.
We now consider a control surface enclosing a blade aggregate but excluding the wake from the trailing edge, such as the blue box in \cref{fig:mirroring_and_solidity_ratio_control_surface}b.
According to Eq.~(4.90) from \citet{Newman2018}, the general expression for the hydrodynamic force (no resistive drag within the potential flow theory) on this aggregate is
\begin{equation}
    \bm{F} = -\rho\frac{\mathrm{d}}{\mathrm{d}t} \iint\limits_{S_B} \phi\bm{n}\mathrm{d}S - \rho \iint\limits_{S_C}\left[\frac{\partial\phi}{\partial n}\nabla\phi - \bm{n}\frac{1}{2}\left(\nabla\phi\cdot\nabla\phi - U^2\right)\right]\mathrm{d}S,
\end{equation}
where $\phi$ is the velocity potential within the control surface $S_C$ and outside the body surface $S_B$, and $\bm{n}$ is the normal vector pointing out of the fluid volume. If the integral on $S_C$ is zero (as in an unbounded fluid), $\bm{F}$ can always be expressed as an inertial or reactive force as \cref{eq:added_mass}.
In our case, since the free surface and the tank bottom are fixed and we assume no flow in the $y$-direction, we have $\partial\phi/\partial n = 0$ on $S_C$.
On the other hand, due to the symmetry of the control volume and the structure about the $xz$-plane, the term $1/2(\nabla\phi\cdot\nabla\phi - U^2)$ contributes nothing in the $y$-direction.
In the $z$-direction, its contribution from the free surface and tank bottom is minimal, as the velocity potential induced by the presence and motion of the blades is small compared to the dominant ambient flow potential. In conclusion, we can still use \cref{eq:added_mass} to calculate the reactive force in a limited fluid.

Since the velocity potential
$\phi$ in a confined domain differs significantly from that in an unbounded fluid, the resulting added mass is also expected to vary accordingly. Our results in \cref{fig:exp_num_drag_reduction} indicate that for some $N$ or $\upbeta$, $C_M=0.1\sim0.6$ gives the best fit, which is quite smaller than unity, the typical value for an individual blade in an infinite fluid domain~\citep{Leclercq2018a,Leclercq2018}.

There are several possible explanations for using cross-sectional $C_M$ values below unity in the numerical model to achieve consistency with the experimental data.
In the flutter regime, flexible blades may encounter a lock-in region similar to that observed in vortex-induced vibrations, where the cross-sectional added mass of cylinders with circular cross-section decreases and can even become negative~\citep{Sarpkaya1978}.
Furthermore, when blades become nearly horizontally aligned in the flutter regime, the downstream part of the blades resides within the wake of the upstream part of the blades convected by the ambient flow. This interaction, which is also known as the $2D+t$ effect~\citep{Alsos2018}, is not included in the hydrodynamic load model.
The model also does not account for asynchronous motion and twisting of blades about their longitudinal axes, as observed in the experiments.
On the other hand, hydrodynamic interaction with neighboring blades is expected to increase the value of $C_M$, potentially above unity.
Our choice of $C_M < 1$, which gives the best fit with the experimental data for most cases, suggests that the $2D+t$ effect, twisting effect, and the possible lock-in behavior must outperform the effect of hydrodynamic interaction between side-by-side blades.

The $C_M$ coefficient is at present only chosen heuristically.
Obtaining physically based values for $C_M$ would advance our model's accuracy significantly.
We see this as a possibility, where the proximity to the neighboring specimen is considered a main element to include.
One can perform forced motions of a single blade with other neighboring blades, in relevant configurations relative to each other, in a dedicated numerical study.
After determining $C_M$ from experiments or numerical simulations, reactive loads on blades can be predicted from their dimensions, which is advantageous in engineering.
However, we leave this as a future study.

\subsection{Limitations}\label{sec:limitations}
There are certain limitations associated with the experiments, as well as the analytical and numerical models.

In the experiments, we measured the total forces acting on both the blade mimics and the supporting structures. To isolate the forces acting solely on the blade mimics, reference measurements were taken without the blades, and the corresponding forces were subtracted. However, because the blades were positioned in the wake of the strut, particularly after the onset of flutter, when the blades aligned more horizontally, interactions between the strut and the blade mimics may have mattered. Consequently, this simple subtraction approach may not fully capture the actual forces on the blades.
To better quantify the influence of the strut in future work, one potential approach would be to systematically vary their dimensions and examine the resulting changes in the measured forces.

Some assumptions made in the analytical and numerical models are not entirely consistent with the experiments, such as the assumption of pure in-plane motion and synchronous motions.

Both theoretical models assume that blade motion is confined to the $xz$-plane. In contrast, the blades in the experiments exhibited three-dimensional motions, including twisting around their longitudinal axes. Blades at the far ends of the strut were sometimes drawn toward the adjacent acrylic plates.

The theoretical models use equivalent thickness and equivalent bending stiffness to represent multiple overlapping blades as a single unit. This approach neglects the interactions between individual blades within an aggregate and their asynchronous movements. In the experiments, adjacent blades could also interact directly through collisions or by hydrodynamic interaction.
While these interactions are not directly captured in the models, they might be incorporated through the coefficients $C_D$ and $C_M$, which can be extracted from future experiments or CFD analysis with side-by-side blades.

\section{Conclusion}\label{sec:conclusion}

This study presents a thorough and systematic investigation, utilizing both experimental and theoretical approaches, into the hydrodynamic drag loads on side-by-side aggregates of flexible blades in a uniform current.
A total of 176 combinations of blade properties (including dimensions and the number of overlapping layers in each aggregate $N$) and current speeds were tested.
The experiments covered two distinct reconfiguration regimes: static and flutter.
We examine four non-dimensional parameters, the Cauchy number, the buoyancy parameter, the mass ratio, and the slenderness parameter, to assess their effects on the bulk drag coefficient for side-by-side blades and the onset of flutter.
When excluding the blockage effect, the bulk drag coefficient remains constant at low Cauchy numbers less than the buoyancy parameter, decreases at moderate Cauchy numbers larger than the buoyancy parameter until flutter occurs, and then transitions to an approximately constant value in the flutter regime at high Cauchy numbers. In the static regime, the bulk drag coefficient is mainly positively correlated with the buoyancy parameter.
The onset of flutter will be delayed to larger Cauchy numbers by increasing the buoyancy parameter, the mass ratio, or the slenderness parameter.
In the flutter regime, a larger mass ratio or slenderness parameter results in a lower bulk drag coefficient.
Notably, the aspect ratio (length to width) is found to have a negligible effect on the bulk drag coefficient or the onset of flutter for side-by-side blades.

Specifically, for a single layer of blades ($N = 1$), the bulk drag coefficient decreases with increasing the Cauchy number until the onset of flutter.
Adding a second layer (from $N=1$ to $N=2$) raises the bulk drag coefficient significantly over the measured range of Cauchy number.
When adding more layers (letting $N = 5$ and $N = 10$), there is a smaller than proportional increase in the bulk drag coefficient and the increase is not consistent over the measured range of Cauchy number.

An analytical model is developed based on the static governing equation of motion for an individual blade, incorporating the concepts of equivalent thickness and bending stiffness to represent an aggregate of blades.
This strategy proves efficient for modeling multiple overlapping blades.
The analytical model accurately predicts drag reduction in the static regime. Furthermore, a numerical model using the same strategy of equivalent thickness and bending stiffness effectively predicts the onset of flutter and drag reduction in both regimes, given certain cross-sectional hydrodynamic coefficients.

Through a review of the various terms in the reactive force model, we identify the term responsible for inducing instability, the term that stabilizes the system, and the term that purely modifies drag reduction. The similarity between the present model and the classic small-amplitude flutter equation for an undamped cantilever beam in axial flow further corroborates these conclusions.
We find that the reactive-force model with only the relative-acceleration term is insufficient to trigger flutter.

The data set presented in this work is expected to be a valuable resource for engineering applications, aiding in the estimation of loads on side-by-side flexible blades. It also serves as a benchmark for theoretical models investigating hydrodynamic loads on such blade arrangements.

\section*{Data Availability}
The supplementary material available at \url{https://doi.org/10.11583/DTU.27180579.v1} includes 1) a Python script for solving the static governing equation~\cref{eq:non_static_ge}, 2) a complete list of all test cases, 3) a lossless version of \cref{fig:exp_photo}, 4) several animations of the numerical simulations, 5) several videos recorded during the experiments, and 6) a Python script for solving the small-amplitude flutter equation~\cref{eq:linear_beam_flutter} for an undamped cantilever beam in axial flow.
The code used for the numerical model is preserved at \url{https://doi.org/10.11583/DTU.24533098.v3}, meanwhile available under the MIT License and developed openly at \url{https://gitlab.gbar.dtu.dk/floatingseaweedfarms/wavevegetationinteraction}.

\section*{Acknowledgments}
Z. Wei and Y. Shao would like to acknowledge the financial support from Innovation Fund Denmark, grant number 3170-00020B.
T. Kristiansen and D. Kristiansen acknowledge partial support from the Research Council of Norway through SFI BLUES, grant number 309281.
Additionally, Z. Wei acknowledges the support and assistance provided by the laboratory staff at NTNU.

\printcredits





\appendix
\section{Uncertainty analysis}\label{app:uncertainty}
\begin{figure}
    \centerline{\includegraphics{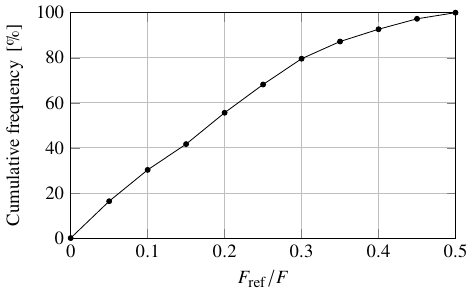}}
    \caption{Contribution from the drag force on the frame $F_{\mathrm{ref}}$ to the total drag force $F$.}
    \label{fig:ref_share_total_loads.pdf}
\end{figure}

There are multiple sources contributing to the uncertainty of $C_{D,\mathrm{bulk}}$.
In our experiments, we measured the total forces and moments acting on both the blades and the holding frame.
To determine the forces on the blades alone, we subtracted the forces on the frame from the total forces.
This calculation assumes there is no interaction between the blades and the strut or the acrylic plates—a factor introducing uncertainty that cannot be directly and accurately quantified.
The proportion of the force on the frame $F_{\mathrm{ref}}$ to the total force $F$ is examined (both computed from the measured $M_y$), as shown in \cref{fig:ref_share_total_loads.pdf}.
In all test cases, $F_{\mathrm{ref}}$ contributed less than \SI{50}{\percent} of $F$.
In nearly \SI{60}{\percent} of the test cases, $F_{\mathrm{ref}}$ contributed less than \SI{20}{\percent} of $F$.
In most cases, the force acting on the blades was significantly greater than that on the frame.
At the higher flow speeds of particular interest, the load contribution from the supporting strut and plates is relatively insignificant.
Although the uncertainty from subtracting the frame's force cannot be precisely quantified, its effect on the bulk drag coefficient is minimal.

However, the uncertainty arising from the variability of the three repetitions of each test condition and from the errors in the material's Young's modulus are quantifiable.
The standard errors of $C_{D,\mathrm{bulk}}$ and $\mathrm{Ca}$ are calculated as follows:
\begin{align}
    \delta C_{D,\mathrm{bulk}} &= \sqrt{\left(\delta F\frac{\partial C_{D,\mathrm{bulk}}}{\partial F}\right)^2 + \left(\delta U\frac{\partial C_{D,\mathrm{bulk}}}{\partial U}\right)^2}, \\
    \delta \mathrm{Ca} &= \sqrt{\left(\delta U\frac{\partial \mathrm{Ca}}{\partial U}\right)^2 + \left(\delta E\frac{\partial \mathrm{Ca}}{\partial E}\right)^2},
\end{align}
where the uncertainties in the horizontal force and velocity, $\delta F$ and $\delta U$ are given by:
\begin{equation}
    \delta F = t_n\sqrt{\frac{1}{n(n-1)}\sum^n_{i=1}\left(F_i - \overline{F}\right)^2}, \delta U = t_n\sqrt{\frac{1}{n(n-1)}\sum^n_{i=1}\left(U_i - \overline{U}\right)^2},\ n = 3,
\end{equation}
and $\overline{F}$ and $\overline{U}$ are the means of the force and velocity measurements $F_i$ and $U_i$, respectively.
The factor $t_n=1.32$ is used due to the small sample size $n=3$, ensuring \SI{68.3}{\percent} confidence intervals~\citep{Student1908}.

\section{Blockage effect}\label{app:blockage_effect}
The analytical model discussed in \cref{sec:analytical_model} considers an individual blade in an infinite fluid domain.
However, in the experiments, the side-by-side blades will block the flow significantly at $\mathrm{Ca}/\mathrm{B} < 1$ when the deflection of blades is small, leading to an overestimation of $C_{D,\mathrm{bulk}}$ as shown in \cref{fig:exp_drag_coefficients}.
We treat the side-by-side blades as a blade with ``holes" or ``gaps" on it.
By considering the physical problem of flexible rectangular blades in a limited fluid domain as illustrated in \cref{fig:mirroring_and_solidity_ratio_control_surface}b, we introduce three additional quantities, the submergence $d_1$ the blade, the distance $d_2 - h_{\mathrm{proj}}$ from the blade tip to the tank bottom, and the gaps $\delta$.
Instead of introducing another three non-dimensional parameters, we follow the approach in studies of perforated plates to include the three quantities in the perforation ratio since the local effective speed is the most important quantity.
There are two factors contributing to the perforation (or inversely the blockage) as shown in \cref{fig:mirroring_and_solidity_ratio_control_surface}b, the gaps above and below the blades as well as the gaps between the blades due to rotations or motion out of phase.
To account for the blockage effect, we define the solidity ratio
\begin{equation}\label{eq:solidity_ratio}
    \mathrm{Sn} = \frac{h_\mathrm{proj}}{h}\left(1 - \frac{\delta}{b}\right),
\end{equation}
where $h_\mathrm{proj}$ is the projected height of the blades in the $yz$-plane, $h=\SI{0.61}{m}$ is water depth, and $\delta$ is the average gap between two aggregates of blades, as illustrated in \cref{fig:mirroring_and_solidity_ratio_control_surface}b.
The gaps between the acrylic plates and the side walls are relatively small compared to the tank width and will be neglected.

For example, when $U=0$ and the blades are hanging vertically downward in the water, $\mathrm{Sn}\approx 5/6$.
As $U$ increases, the blades become more aligned with the horizontal direction, causing $\mathrm{Sn}$ to decrease.
The local effective velocity at the blades (or over the blade span in $x$-direction) can be approximated by $U^{\mathrm{local}}=U/(1 - \mathrm{Sn})$,
where the factor $1/(1 - \mathrm{Sn})$ accounts for the local speed-up of the flow as it passes around the blades, due to mass conservation.

The local $\mathrm{Ca}^{\mathrm{local}}$ can then be expressed as $\mathrm{Ca}^{\mathrm{local}}=\mathrm{Ca}/(1 - \mathrm{Sn})^2$ from $\mathrm{Ca}^{\mathrm{local}}/\mathrm{Ca} = (U^{\mathrm{local}}/U)^2$.
Using this relation, we adjust the analytical solution (represented by solid curves in the \cref{fig:exp_ana_drag_reduction}) as follows:
\begin{enumerate}
    \item~For a given $\mathrm{Ca}$, solve \cref{eq:non_static_ge} to obtain $h_{\mathrm{proj}}=l\int^1_0 \cos\theta\mathrm{d}\tilde{s}$ and $\mathcal{R}^{\mathrm{ana,l}}$, and compute $\mathrm{Sn}$ using \cref{eq:solidity_ratio}.
    \item~Update $\mathrm{Ca}^{\mathrm{local}}=\mathrm{Ca}/(1 - \mathrm{Sn})^2$.
    \item~With the updated $\mathrm{Ca}^{\mathrm{local}}$, solve \cref{eq:non_static_ge} again to obtain updated $h_{\mathrm{proj}}$ and $\mathcal{R}^{\mathrm{ana,l}}$, and compute $\mathrm{Sn}$.
    \item~Repeat Steps (ii) and (iii) until $\mathcal{R}^{\mathrm{ana,l}}$ converges.
    \item~Calculate the corrected $\mathcal{R}^{\mathrm{CWT}} = \mathcal{R}(\mathrm{Ca}^{\mathrm{local}}) \mathrm{Ca}^{\mathrm{local}} / \mathrm{Ca} $. \\
\end{enumerate}
Through this iterative approach, we determine $\mathrm{Ca}$ using the known upstream velocity and compute the corresponding $\mathcal{R}$ at this velocity, including the effects of blockage. This allows us to compare the analytical solution with the experimental data.
As shown in \cref{fig:exp_ana_drag_reduction}, the blockage effect is significant only at low $\mathrm{Ca}$.
At high $\mathrm{Ca}$, the blades align with the flow direction, and the solidity ratio approaches zero. When accounting for the blockage effect, the analytical predictions at small $\mathrm{Ca}$ generally align with the experimental data, although they are slightly lower.

\bibliographystyle{cas-model2-names}


\begin{thebibliography}{45}
\expandafter\ifx\csname natexlab\endcsname\relax\def\natexlab#1{#1}\fi
\providecommand{\url}[1]{\texttt{#1}}
\providecommand{\href}[2]{#2}
\providecommand{\path}[1]{#1}
\providecommand{\DOIprefix}{doi:}
\providecommand{\ArXivprefix}{arXiv:}
\providecommand{\URLprefix}{URL: }
\providecommand{\Pubmedprefix}{pmid:}
\providecommand{\doi}[1]{\href{http://dx.doi.org/#1}{\path{#1}}}
\providecommand{\Pubmed}[1]{\href{pmid:#1}{\path{#1}}}
\providecommand{\bibinfo}[2]{#2}
\ifx\xfnm\relax \def\xfnm[#1]{\unskip,\space#1}\fi
\bibitem[{Alsos and Faltinsen(2018)}]{Alsos2018}
\bibinfo{author}{Alsos, H.S.}, \bibinfo{author}{Faltinsen, O.M.}, \bibinfo{year}{2018}.
\newblock \bibinfo{title}{{3D} motion dynamics of axisymmetric bodies falling through water}.
\newblock \bibinfo{journal}{Ocean Engineering} \bibinfo{volume}{169}, \bibinfo{pages}{442--456}.
\newblock \URLprefix \url{https://www.sciencedirect.com/science/article/pii/S0029801818315890}, \DOIprefix\doi{10.1016/j.oceaneng.2018.08.033}.
\bibitem[{Baskaran et~al.(2023)Baskaran, Hutin and Mulleners}]{Baskaran2023}
\bibinfo{author}{Baskaran, M.}, \bibinfo{author}{Hutin, L.}, \bibinfo{author}{Mulleners, K.}, \bibinfo{year}{2023}.
\newblock \bibinfo{title}{Reconfiguring it out: {How} flexible structures interact with fluid flows}.
\newblock \bibinfo{journal}{Physical Review Fluids} \bibinfo{volume}{8}, \bibinfo{pages}{110509}.
\newblock \URLprefix \url{https://link.aps.org/doi/10.1103/PhysRevFluids.8.110509}, \DOIprefix\doi{10.1103/PhysRevFluids.8.110509}.
\bibitem[{Boukor et~al.(2024)Boukor, Choimet, Laurendeau and Gosselin}]{Boukor2024}
\bibinfo{author}{Boukor, M.}, \bibinfo{author}{Choimet, A.}, \bibinfo{author}{Laurendeau, E.}, \bibinfo{author}{Gosselin, F.P.}, \bibinfo{year}{2024}.
\newblock \bibinfo{title}{Flutter limitation of drag reduction by elastic reconfiguration}.
\newblock \bibinfo{journal}{Physics of Fluids} \bibinfo{volume}{36}, \bibinfo{pages}{021915}.
\newblock \URLprefix \url{https://doi.org/10.1063/5.0193649}, \DOIprefix\doi{10.1063/5.0193649}.
\bibitem[{Buck and Buchholz(2005)}]{Buck2005}
\bibinfo{author}{Buck, B.H.}, \bibinfo{author}{Buchholz, C.M.}, \bibinfo{year}{2005}.
\newblock \bibinfo{title}{{Response} of offshore cultivated {Laminaria} saccharina to hydrodynamic forcing in the {North} {Sea}}.
\newblock \bibinfo{journal}{Aquaculture} \bibinfo{volume}{250}, \bibinfo{pages}{674--691}.
\newblock \DOIprefix\doi{10.1016/j.aquaculture.2005.04.062}.
\bibitem[{Campbell et~al.(2019)Campbell, Macleod, Sahlmann, Neves, Funderud, Øverland, Hughes and Stanley}]{Campbell2019}
\bibinfo{author}{Campbell, I.}, \bibinfo{author}{Macleod, A.}, \bibinfo{author}{Sahlmann, C.}, \bibinfo{author}{Neves, L.}, \bibinfo{author}{Funderud, J.}, \bibinfo{author}{Øverland, M.}, \bibinfo{author}{Hughes, A.D.}, \bibinfo{author}{Stanley, M.}, \bibinfo{year}{2019}.
\newblock \bibinfo{title}{{The} {Environmental} {Risks} {Associated} {With} the {Development} of {Seaweed} {Farming} in {Europe} - {Prioritizing} {Key} {Knowledge} {Gaps}}.
\newblock \bibinfo{journal}{Frontiers in Marine Science} \bibinfo{volume}{6}.
\newblock \DOIprefix\doi{10.3389/fmars.2019.00107}.
\bibitem[{Candelier et~al.(2011)Candelier, Boyer and Leroyer}]{Candelier2011}
\bibinfo{author}{Candelier, F.}, \bibinfo{author}{Boyer, F.}, \bibinfo{author}{Leroyer, A.}, \bibinfo{year}{2011}.
\newblock \bibinfo{title}{Three-dimensional extension of {Lighthill}'s large-amplitude elongated-body theory of fish locomotion}.
\newblock \bibinfo{journal}{Journal of Fluid Mechanics} \bibinfo{volume}{674}, \bibinfo{pages}{196--226}.
\newblock \URLprefix \url{https://www.cambridge.org/core/journals/journal-of-fluid-mechanics/article/threedimensional-extension-of-lighthills-largeamplitude-elongatedbody-theory-of-fish-locomotion/6671261A9407EE44D0700BEAC9CB7C73}, \DOIprefix\doi{10.1017/S002211201000649X}.
\bibitem[{Carrington(1990)}]{Carrington1990}
\bibinfo{author}{Carrington, E.}, \bibinfo{year}{1990}.
\newblock \bibinfo{title}{Drag and dislodgment of an intertidal macroalga: consequences of morphological variation in \textit{{Mastocarpus} papillatus} {Kützing}}.
\newblock \bibinfo{journal}{Journal of Experimental Marine Biology and Ecology} \bibinfo{volume}{139}, \bibinfo{pages}{185--200}.
\newblock \URLprefix \url{https://www.sciencedirect.com/science/article/pii/0022098190901464}, \DOIprefix\doi{10.1016/0022-0981(90)90146-4}.
\bibitem[{Eloy et~al.(2011)Eloy, Kofman and Schouveiler}]{Eloy2011}
\bibinfo{author}{Eloy, C.}, \bibinfo{author}{Kofman, N.}, \bibinfo{author}{Schouveiler, L.}, \bibinfo{year}{2011}.
\newblock \bibinfo{title}{The origin of hysteresis in the flag instability}.
\newblock \bibinfo{journal}{Journal of Fluid Mechanics} \bibinfo{volume}{691}, \bibinfo{pages}{583--593}.
\newblock \DOIprefix\doi{10.1017/jfm.2011.494}.
\bibitem[{Fredriksson et~al.(2020)Fredriksson, Dewhurst, Drach, Beaver, Gelais, Johndrow and Costa-Pierce}]{Fredriksson2020}
\bibinfo{author}{Fredriksson, D.W.}, \bibinfo{author}{Dewhurst, T.}, \bibinfo{author}{Drach, A.}, \bibinfo{author}{Beaver, W.}, \bibinfo{author}{Gelais, A.T.S.}, \bibinfo{author}{Johndrow, K.}, \bibinfo{author}{Costa-Pierce, B.A.}, \bibinfo{year}{2020}.
\newblock \bibinfo{title}{Hydrodynamic characteristics of a full-scale kelp model for aquaculture applications}.
\newblock \bibinfo{journal}{Aquacultural Engineering} \bibinfo{volume}{90}, \bibinfo{pages}{102086}.
\newblock \DOIprefix\doi{10.1016/j.aquaeng.2020.102086}.
\bibitem[{{GOSSELIN} et~al.(2010){GOSSELIN}, {de LANGRE} and {MACHADO-ALMEIDA}}]{Gosselin2010}
\bibinfo{author}{{GOSSELIN}, F.}, \bibinfo{author}{{de LANGRE}, E.}, \bibinfo{author}{{MACHADO-ALMEIDA}, B.}, \bibinfo{year}{2010}.
\newblock \bibinfo{title}{Drag reduction of flexible plates by reconfiguration}.
\newblock \bibinfo{journal}{Journal of Fluid Mechanics} \bibinfo{volume}{650}, \bibinfo{pages}{319--341}.
\newblock \DOIprefix\doi{10.1017/s0022112009993673}.
\bibitem[{Grebe et~al.(2019)Grebe, Byron, Gelais, Kotowicz and Olson}]{Grebe2019}
\bibinfo{author}{Grebe, G.S.}, \bibinfo{author}{Byron, C.J.}, \bibinfo{author}{Gelais, A.S.}, \bibinfo{author}{Kotowicz, D.M.}, \bibinfo{author}{Olson, T.K.}, \bibinfo{year}{2019}.
\newblock \bibinfo{title}{An ecosystem approach to kelp aquaculture in the {Americas} and {Europe}}.
\newblock \bibinfo{journal}{Aquaculture Reports} \bibinfo{volume}{15}, \bibinfo{pages}{100215}.
\newblock \URLprefix \url{https://www.sciencedirect.com/science/article/pii/S2352513419300134}, \DOIprefix\doi{10.1016/j.aqrep.2019.100215}.
\bibitem[{Henriquez and Barrero-Gil(2014)}]{Henriquez2014}
\bibinfo{author}{Henriquez, S.}, \bibinfo{author}{Barrero-Gil, A.}, \bibinfo{year}{2014}.
\newblock \bibinfo{title}{Reconfiguration of flexible plates in sheared flow}.
\newblock \bibinfo{journal}{Mechanics Research Communications} \bibinfo{volume}{62}, \bibinfo{pages}{1--4}.
\newblock \URLprefix \url{https://www.sciencedirect.com/science/article/pii/S0093641314001062}, \DOIprefix\doi{10.1016/j.mechrescom.2014.08.001}.
\bibitem[{Jacobsen et~al.(2019)Jacobsen, Bakker, Uijttewaal and Uittenbogaard}]{Jacobsen2019a}
\bibinfo{author}{Jacobsen, N.G.}, \bibinfo{author}{Bakker, W.}, \bibinfo{author}{Uijttewaal, W.S.J.}, \bibinfo{author}{Uittenbogaard, R.}, \bibinfo{year}{2019}.
\newblock \bibinfo{title}{Experimental investigation of the wave-induced motion of and force distribution along a flexible stem}.
\newblock \bibinfo{journal}{Journal of Fluid Mechanics} \bibinfo{volume}{880}, \bibinfo{pages}{1036--1069}.
\newblock \DOIprefix\doi{10.1017/jfm.2019.739}.
\bibitem[{Johnson(2001)}]{Johnson2001}
\bibinfo{author}{Johnson, A.S.}, \bibinfo{year}{2001}.
\newblock \bibinfo{title}{{Drag}, {Drafting}, and {Mechanical} {Interactions} in {Canopies} of the {Red} {Alga} {Chondrus} crispus}.
\newblock \bibinfo{journal}{The Biological Bulletin} \bibinfo{volume}{201}, \bibinfo{pages}{126--135}.
\newblock \DOIprefix\doi{10.2307/1543328}.
\bibitem[{Kobayashi et~al.(1993)Kobayashi, Raichle and Asano}]{Kobayashi1993}
\bibinfo{author}{Kobayashi, N.}, \bibinfo{author}{Raichle, A.W.}, \bibinfo{author}{Asano, T.}, \bibinfo{year}{1993}.
\newblock \bibinfo{title}{{Wave} {Attenuation} by {Vegetation}}.
\newblock \bibinfo{journal}{Journal of Waterway, Port, Coastal, and Ocean Engineering} \bibinfo{volume}{119}, \bibinfo{pages}{30--48}.
\newblock \DOIprefix\doi{10.1061/(asce)0733-950x(1993)119:1(30)}.
\bibitem[{Kristiansen and Faltinsen(2012)}]{Kristiansen2012}
\bibinfo{author}{Kristiansen, T.}, \bibinfo{author}{Faltinsen, O.M.}, \bibinfo{year}{2012}.
\newblock \bibinfo{title}{Modelling of current loads on aquaculture net cages}.
\newblock \bibinfo{journal}{Journal of Fluids and Structures} \bibinfo{volume}{34}, \bibinfo{pages}{218--235}.
\newblock \DOIprefix\doi{10.1016/j.jfluidstructs.2012.04.001}.
\bibitem[{Kristiansen and Faltinsen(2015)}]{Kristiansen2015}
\bibinfo{author}{Kristiansen, T.}, \bibinfo{author}{Faltinsen, O.M.}, \bibinfo{year}{2015}.
\newblock \bibinfo{title}{Experimental and numerical study of an aquaculture net cage with floater in waves and current}.
\newblock \bibinfo{journal}{Journal of Fluids and Structures} \bibinfo{volume}{54}, \bibinfo{pages}{1--26}.
\newblock \DOIprefix\doi{10.1016/j.jfluidstructs.2014.08.015}.
\bibitem[{Leclercq and de~Langre(2018)}]{Leclercq2018}
\bibinfo{author}{Leclercq, T.}, \bibinfo{author}{de~Langre, E.}, \bibinfo{year}{2018}.
\newblock \bibinfo{title}{Reconfiguration of elastic blades in oscillatory~flow}.
\newblock \bibinfo{journal}{Journal of Fluid Mechanics} \bibinfo{volume}{838}, \bibinfo{pages}{606--630}.
\newblock \DOIprefix\doi{10.1017/jfm.2017.910}.
\bibitem[{Leclercq et~al.(2018)Leclercq, Peake and de~Langre}]{Leclercq2018a}
\bibinfo{author}{Leclercq, T.}, \bibinfo{author}{Peake, N.}, \bibinfo{author}{de~Langre, E.}, \bibinfo{year}{2018}.
\newblock \bibinfo{title}{Does flutter prevent drag reduction by reconfiguration?}
\newblock \bibinfo{journal}{Proceedings of the Royal Society A: Mathematical, Physical and Engineering Sciences} \bibinfo{volume}{474}, \bibinfo{pages}{20170678}.
\newblock \URLprefix \url{https://royalsocietypublishing.org/doi/full/10.1098/rspa.2017.0678}, \DOIprefix\doi{10.1098/rspa.2017.0678}.
\bibitem[{Lei et~al.(2021)Lei, Fan, Angera, Liu and Nepf}]{Lei2021}
\bibinfo{author}{Lei, J.}, \bibinfo{author}{Fan, D.}, \bibinfo{author}{Angera, A.}, \bibinfo{author}{Liu, Y.}, \bibinfo{author}{Nepf, H.}, \bibinfo{year}{2021}.
\newblock \bibinfo{title}{Drag force and reconfiguration of cultivated {Saccharina} latissima in current}.
\newblock \bibinfo{journal}{Aquacultural Engineering} \bibinfo{volume}{94}, \bibinfo{pages}{102169}.
\newblock \URLprefix \url{https://www.sciencedirect.com/science/article/pii/S014486092100025X}, \DOIprefix\doi{10.1016/j.aquaeng.2021.102169}.
\bibitem[{Lei and Nepf(2019a)}]{Lei2019}
\bibinfo{author}{Lei, J.}, \bibinfo{author}{Nepf, H.}, \bibinfo{year}{2019}a.
\newblock \bibinfo{title}{Blade dynamics in combined waves and current}.
\newblock \bibinfo{journal}{Journal of Fluids and Structures} \bibinfo{volume}{87}, \bibinfo{pages}{137--149}.
\newblock \DOIprefix\doi{10.1016/j.jfluidstructs.2019.03.020}.
\bibitem[{Lei and Nepf(2019b)}]{Lei2019a}
\bibinfo{author}{Lei, J.}, \bibinfo{author}{Nepf, H.}, \bibinfo{year}{2019}b.
\newblock \bibinfo{title}{Wave damping by flexible vegetation: {Connecting} individual blade dynamics to the meadow scale}.
\newblock \bibinfo{journal}{Coastal Engineering} \bibinfo{volume}{147}, \bibinfo{pages}{138--148}.
\newblock \URLprefix \url{https://www.sciencedirect.com/science/article/pii/S0378383918300905}, \DOIprefix\doi{10.1016/j.coastaleng.2019.01.008}.
\bibitem[{Lighthill(1971)}]{Lighthill1971}
\bibinfo{author}{Lighthill, M.J.}, \bibinfo{year}{1971}.
\newblock \bibinfo{title}{Large-amplitude elongated-body theory of fish locomotion}.
\newblock \bibinfo{journal}{Proceedings of the Royal Society of London. Series B. Biological Sciences} \bibinfo{volume}{179}, \bibinfo{pages}{125--138}.
\newblock \URLprefix \url{https://royalsocietypublishing.org/doi/10.1098/rspb.1971.0085}, \DOIprefix\doi{10.1098/rspb.1971.0085}.
\bibitem[{Luhar et~al.(2010)Luhar, Coutu, Infantes, Fox and Nepf}]{Luhar2010}
\bibinfo{author}{Luhar, M.}, \bibinfo{author}{Coutu, S.}, \bibinfo{author}{Infantes, E.}, \bibinfo{author}{Fox, S.}, \bibinfo{author}{Nepf, H.}, \bibinfo{year}{2010}.
\newblock \bibinfo{title}{Wave‐induced velocities inside a model seagrass bed}.
\newblock \bibinfo{journal}{Journal of Geophysical Research: Oceans} \bibinfo{volume}{115}.
\newblock \DOIprefix\doi{10.1029/2010jc006345}.
\bibitem[{Luhar et~al.(2017)Luhar, Infantes and Nepf}]{Luhar2017}
\bibinfo{author}{Luhar, M.}, \bibinfo{author}{Infantes, E.}, \bibinfo{author}{Nepf, H.}, \bibinfo{year}{2017}.
\newblock \bibinfo{title}{Seagrass blade motion under waves and its impact on wave decay}.
\newblock \bibinfo{journal}{Journal of Geophysical Research: Oceans} \bibinfo{volume}{122}, \bibinfo{pages}{3736--3752}.
\newblock \URLprefix \url{https://onlinelibrary.wiley.com/doi/abs/10.1002/2017JC012731}, \DOIprefix\doi{10.1002/2017JC012731}.
\bibitem[{Luhar and Nepf(2016)}]{Luhar2016}
\bibinfo{author}{Luhar, M.}, \bibinfo{author}{Nepf, H.}, \bibinfo{year}{2016}.
\newblock \bibinfo{title}{Wave-induced dynamics of flexible blades}.
\newblock \bibinfo{journal}{Journal of Fluids and Structures} \bibinfo{volume}{61}, \bibinfo{pages}{20--41}.
\newblock \DOIprefix\doi{10.1016/j.jfluidstructs.2015.11.007}.
\bibitem[{Luhar and Nepf(2011)}]{Luhar2011}
\bibinfo{author}{Luhar, M.}, \bibinfo{author}{Nepf, H.M.}, \bibinfo{year}{2011}.
\newblock \bibinfo{title}{Flow-induced reconfiguration of buoyant and flexible aquatic vegetation}.
\newblock \bibinfo{journal}{Limnology and Oceanography} \bibinfo{volume}{56}, \bibinfo{pages}{2003--2017}.
\newblock \DOIprefix\doi{10.4319/lo.2011.56.6.2003}.
\bibitem[{Marichal(2003)}]{Marichal2003}
\bibinfo{author}{Marichal, D.}, \bibinfo{year}{2003}.
\newblock \bibinfo{title}{Cod-end numerical study}, in: \bibinfo{booktitle}{Proceedings of the 3rd International Conference on Hydroelasticity in Marine Technology}, \bibinfo{address}{Oxford, United Kingdom}.
\bibitem[{Neufeldt et~al.(2025)Neufeldt, Windt, Buck, Heasman, Hildebrandt and Goseberg}]{Neufeldt2025}
\bibinfo{author}{Neufeldt, H.}, \bibinfo{author}{Windt, C.}, \bibinfo{author}{Buck, B.H.}, \bibinfo{author}{Heasman, K.}, \bibinfo{author}{Hildebrandt, A.}, \bibinfo{author}{Goseberg, N.}, \bibinfo{year}{2025}.
\newblock \bibinfo{title}{Physical and numerical modeling of seaweed in oceanic waters}.
\newblock \bibinfo{journal}{Aquacultural Engineering} \bibinfo{volume}{110}, \bibinfo{pages}{102528}.
\newblock \DOIprefix\doi{10.1016/j.aquaeng.2025.102528}.
\bibitem[{Newman(2018)}]{Newman2018}
\bibinfo{author}{Newman, J.N.}, \bibinfo{year}{2018}.
\newblock \bibinfo{title}{Marine hydrodynamics}.
\newblock \bibinfo{publisher}{The MIT press}.
\bibitem[{Paidoussis(1998)}]{Paidoussis1998}
\bibinfo{author}{Paidoussis, M.P.}, \bibinfo{year}{1998}.
\newblock \bibinfo{title}{Fluid-{Structure} {Interactions}: {Slender} {Structures} and {Axial} {Flow}}.
\newblock \bibinfo{publisher}{Academic Press}.
\bibitem[{Peirce(1930)}]{Peirce1930}
\bibinfo{author}{Peirce, F.T.}, \bibinfo{year}{1930}.
\newblock \bibinfo{title}{26—the “{Handle}” of {Cloth} as a {Measurable} {Quantity}}.
\newblock \bibinfo{journal}{Journal of the Textile Institute Transactions} \bibinfo{volume}{21}, \bibinfo{pages}{T377--T416}.
\newblock \URLprefix \url{https://doi.org/10.1080/19447023008661529}, \DOIprefix\doi{10.1080/19447023008661529}.
\bibitem[{Sarpkaya(1978)}]{Sarpkaya1978}
\bibinfo{author}{Sarpkaya, T.}, \bibinfo{year}{1978}.
\newblock \bibinfo{title}{Fluid {Forces} on {Oscillating} {Cylinders}}.
\newblock \bibinfo{journal}{Journal of the Waterway, Port, Coastal and Ocean Division} \bibinfo{volume}{104}, \bibinfo{pages}{275--290}.
\newblock \URLprefix \url{https://ascelibrary.org/doi/10.1061/JWPCDX.0000101}, \DOIprefix\doi{10.1061/JWPCDX.0000101}.
\bibitem[{Schaefer and Nepf(2024)}]{Schaefer2024}
\bibinfo{author}{Schaefer, R.}, \bibinfo{author}{Nepf, H.}, \bibinfo{year}{2024}.
\newblock \bibinfo{title}{Movement of and drag force on slender flat plates in an array exposed to combinations of unidirectional and oscillatory flow}.
\newblock \bibinfo{journal}{Journal of Fluids and Structures} \bibinfo{volume}{124}, \bibinfo{pages}{104044}.
\newblock \URLprefix \url{https://www.sciencedirect.com/science/article/pii/S0889974623002128}, \DOIprefix\doi{10.1016/j.jfluidstructs.2023.104044}.
\bibitem[{Singh et~al.(2012)Singh, Michelin and de~Langre}]{Singh2012}
\bibinfo{author}{Singh, K.}, \bibinfo{author}{Michelin, S.}, \bibinfo{author}{de~Langre, E.}, \bibinfo{year}{2012}.
\newblock \bibinfo{title}{Energy harvesting from axial fluid-elastic instabilities of a cylinder}.
\newblock \bibinfo{journal}{Journal of Fluids and Structures} \bibinfo{volume}{30}, \bibinfo{pages}{159--172}.
\newblock \URLprefix \url{https://www.sciencedirect.com/science/article/pii/S0889974612000321}, \DOIprefix\doi{10.1016/j.jfluidstructs.2012.01.008}.
\bibitem[{{Student}(1908)}]{Student1908}
\bibinfo{author}{{Student}}, \bibinfo{year}{1908}.
\newblock \bibinfo{title}{The {Probable} {Error} of a {Mean}}.
\newblock \bibinfo{journal}{Biometrika} \bibinfo{volume}{6}, \bibinfo{pages}{1--25}.
\newblock \URLprefix \url{https://www.jstor.org/stable/2331554}, \DOIprefix\doi{10.2307/2331554}.
\bibitem[{Sun et~al.(2024)Sun, You and Lei}]{Sun2024}
\bibinfo{author}{Sun, H.}, \bibinfo{author}{You, Y.}, \bibinfo{author}{Lei, J.}, \bibinfo{year}{2024}.
\newblock \bibinfo{title}{Deflection and drag on flexible marine structures in steady currents and internal solitary waves}.
\newblock \bibinfo{journal}{Physics of Fluids} \bibinfo{volume}{36}, \bibinfo{pages}{106603}.
\newblock \URLprefix \url{https://doi.org/10.1063/5.0227279}, \DOIprefix\doi{10.1063/5.0227279}.
\bibitem[{Taylor(1952)}]{Taylor1952}
\bibinfo{author}{Taylor, G.I.}, \bibinfo{year}{1952}.
\newblock \bibinfo{title}{Analysis of the swimming of long and narrow animals}.
\newblock \bibinfo{journal}{Proceedings of the Royal Society of London. Series A. Mathematical and Physical Sciences} \bibinfo{volume}{214}, \bibinfo{pages}{158--183}.
\newblock \DOIprefix\doi{10.1098/rspa.1952.0159}.
\bibitem[{Vettori et~al.(2024)Vettori, Pezzutto, Bouma, Shahmohammadi and Manes}]{Vettori2024}
\bibinfo{author}{Vettori, D.}, \bibinfo{author}{Pezzutto, P.}, \bibinfo{author}{Bouma, T.J.}, \bibinfo{author}{Shahmohammadi, A.}, \bibinfo{author}{Manes, C.}, \bibinfo{year}{2024}.
\newblock \bibinfo{title}{On the wave attenuation properties of seagrass meadows}.
\newblock \bibinfo{journal}{Coastal Engineering} \bibinfo{volume}{189}, \bibinfo{pages}{104472}.
\newblock \URLprefix \url{https://www.sciencedirect.com/science/article/pii/S0378383924000206}, \DOIprefix\doi{10.1016/j.coastaleng.2024.104472}.
\bibitem[{Vogel(1989)}]{Vogel1989}
\bibinfo{author}{Vogel, S.}, \bibinfo{year}{1989}.
\newblock \bibinfo{title}{Drag and {Reconfiguration} of {Broad} {Leaves} in {High} {Winds}}.
\newblock \bibinfo{journal}{Journal of Experimental Botany} \bibinfo{volume}{40}, \bibinfo{pages}{941--948}.
\newblock \URLprefix \url{https://www.jstor.org/stable/23692362}.
\bibitem[{Vogel(1996)}]{Vogel1996}
\bibinfo{author}{Vogel, S.}, \bibinfo{year}{1996}.
\newblock \bibinfo{title}{Life in {Moving} {Fluids}: {The} {Physical} {Biology} of {Flow}}.
\newblock \bibinfo{edition}{2nd revised edition} ed., \bibinfo{publisher}{Princeton University Press}, \bibinfo{address}{Princeton, NJ}.
\bibitem[{Wei et~al.(2024)Wei, Shao, Kristiansen and Kristiansen}]{Wei2024a}
\bibinfo{author}{Wei, Z.}, \bibinfo{author}{Shao, Y.}, \bibinfo{author}{Kristiansen, T.}, \bibinfo{author}{Kristiansen, D.}, \bibinfo{year}{2024}.
\newblock \bibinfo{title}{An efficient numerical solver for highly compliant slender structures in waves: {Application} to marine vegetation}.
\newblock \bibinfo{journal}{Journal of Fluids and Structures} \bibinfo{volume}{129}, \bibinfo{pages}{104170}.
\newblock \URLprefix \url{https://www.sciencedirect.com/science/article/pii/S0889974624001051}, \DOIprefix\doi{10.1016/j.jfluidstructs.2024.104170}.
\bibitem[{Whittaker et~al.(2015)Whittaker, Wilson and Aberle}]{Whittaker2015}
\bibinfo{author}{Whittaker, P.}, \bibinfo{author}{Wilson, C.A.M.E.}, \bibinfo{author}{Aberle, J.}, \bibinfo{year}{2015}.
\newblock \bibinfo{title}{An improved {Cauchy} number approach for predicting the drag and reconfiguration of flexible vegetation}.
\newblock \bibinfo{journal}{Advances in Water Resources} \bibinfo{volume}{83}, \bibinfo{pages}{28--35}.
\newblock \URLprefix \url{https://www.sciencedirect.com/science/article/pii/S0309170815000986}, \DOIprefix\doi{10.1016/j.advwatres.2015.05.005}.
\bibitem[{Zhang and Nepf(2020)}]{Zhang2020}
\bibinfo{author}{Zhang, X.}, \bibinfo{author}{Nepf, H.}, \bibinfo{year}{2020}.
\newblock \bibinfo{title}{Flow-induced reconfiguration of aquatic plants, including the impact of leaf sheltering}.
\newblock \bibinfo{journal}{Limnology and Oceanography} \bibinfo{volume}{65}, \bibinfo{pages}{2697--2712}.
\newblock \URLprefix \url{https://onlinelibrary.wiley.com/doi/abs/10.1002/lno.11542}, \DOIprefix\doi{10.1002/lno.11542}.
\bibitem[{Zhu et~al.(2021)Zhu, Lei, Huguenard and Fredriksson}]{Zhu2021}
\bibinfo{author}{Zhu, L.}, \bibinfo{author}{Lei, J.}, \bibinfo{author}{Huguenard, K.}, \bibinfo{author}{Fredriksson, D.W.}, \bibinfo{year}{2021}.
\newblock \bibinfo{title}{{Wave} attenuation by suspended canopies with cultivated kelp ({Saccharina} latissima)}.
\newblock \bibinfo{journal}{Coastal Engineering} \bibinfo{volume}{168}, \bibinfo{pages}{103947}.
\newblock \DOIprefix\doi{10.1016/j.coastaleng.2021.103947}.

\end{thebibliography}



\end{document}